\documentclass[a4paper,12pt]{article}
\usepackage{lmodern}
\usepackage[font={footnotesize,doublespacing}]{caption}
\usepackage{microtype} 
\usepackage{fancyhdr}
\usepackage{setspace}
\usepackage{etex}
\usepackage{hyperref}
\hypersetup{
    colorlinks=true,
    linkcolor=blue,
    citecolor=blue,
    filecolor=blue,      
    urlcolor=cyan,
    }
\usepackage[a4paper,margin=1in]{geometry}

\usepackage{amsmath, amssymb, amsthm, mathtools, braket}
\usepackage{authblk}
\usepackage{booktabs}
\usepackage{listings}

\usepackage{graphicx}
\usepackage{caption}
\usepackage{subcaption}
\usepackage[dvipsnames]{xcolor}

\usepackage[numbers,sort&compress]{natbib}
\bibpunct[, ]{[}{]}{,}{n}{,}{,}

\usepackage{doi}
\usepackage{url}
\renewcommand{\url}[1]{\href{#1}{#1}}

\title{\textbf{The Quantum Measurement Problem: A Review of Recent Trends}}

\author[1]{Anderson A. Tomaz \thanks{Corresponding author: \href{mailto:anderson.alves-tomaz@univ-amu.fr}{anderson.alves-tomaz@univ-amu.fr}}}
\author[1]{Rafael S. Mattos}
\author[1,2]{Mario Barbatti \thanks{Corresponding author: \href{mailto:mario.barbatti@univ-amu.fr}{mario.barbatti@univ-amu.fr} | \href{https://www.barbatti.org}{www.barbatti.org}}}
\affil[1]{\small\textit{\fontfamily{lmss}\selectfont Aix Marseille University, CNRS, ICR, Marseille, France}}
\affil[2]{\small \textit{\fontfamily{lmss}\selectfont Institut Universitaire de France, 75231 Paris, France}}

\date{}

\begin{document}

\fontfamily{lmss}\selectfont

\maketitle

\begin{abstract}
\noindent Left on its own, a quantum state evolves deterministically under the Schr\"odinger equation, forming superpositions. Upon measurement, however, a stochastic process governed by the Born rule collapses it to a single outcome. This dual evolution of quantum states---the core of the Measurement Problem---has puzzled physicists and philosophers for nearly a century. Yet, amid the cacophony of competing interpretations, the problem today is not as impenetrable as it once seemed. This paper reviews the current status of the Measurement Problem, distinguishing between what is well understood and what remains unresolved. We examine key theoretical approaches, including decoherence, many-worlds interpretation, objective collapse theories, hidden-variable theories, dualistic approaches, deterministic models, and epistemic interpretations. To make these discussions accessible to a broader audience, we also reference curated online resources that provide high-quality introductions to central concepts.
\end{abstract}

Keywords: decoherence; measurement problem; quantum state collapse; interpretations of quantum mechanics 

\tableofcontents

\section{Introduction}

Quantum mechanics, as we learn in college, posits that an isolated system evolves unitarily and deterministically following the Schr\"odinger equation. When we measure the system, the Born rule determines the output probabilities. The quantum state collapses in a non-unitary, stochastic way, so the same outcome is obtained if the measurement is repeated. Despite its awkward double rule of evolution, this seemingly simple framework enabled comprehension of nature and technological advances to a level that our species has never seen before.

Nevertheless, let us ask what “measure the system” means. We will not find an answer in the standard formulation of quantum mechanics. This innocent question unravels a disturbing thread of conceptual problems, which are surprisingly difficult to address experimentally and may have dramatic philosophical consequences for our conception of the world.

The Measurement Problem has been haunting physicists and philosophers of science for almost as long as the century since quantum mechanics was proposed. We can take as an anecdotal example the World Science Festival held in the first semester of 2024, a science popularization event hosted by Brian Greene. Greene \href{https://www.youtube.com/watch?v=twY2q1F-ciI}{asked} three quantum mechanics specialists about the Measurement Problem and got three entirely distinct answers. The entry \href{https://en.wikipedia.org/wiki/Interpretations_of_quantum_mechanics}{Interpretations of Quantum Mechanics} in Wikipedia contains thirteen distinct classes of theories and interpretations, and it does not even account for some recent proposals.

The cacophony of proposed solutions to the Measurement Problem---including voices from parts of the physics community that question whether the problem even exists---is disturbing. It hurts our conception of science as delivering a consistent description of reality. Why, after a century of quantum mechanics, do we still have no consensus on its most fundamental process? What do we truly understand about the Measurement Problem?

Any newcomer trying to get an answer to these questions faces a formidable challenge: the Measurement Problem research is spread throughout countless isolated sub-communities. These groups do not talk to each other, making it exceptionally difficult to build an overview of what's going on. The large number of single-author papers doesn’t help either. These papers raise hypotheses and propose theories and tests. However, most don’t resonate within a mature community and only add to the cacophony.

As newcomers to the field ourselves---we just published our first contribution recently \cite{Tomaz2024}---we felt firsthand all these issues. We tackled this challenge with extensive reading, discussion, and consultation with specialists on specific topics. We explored all corners of the field rather than focusing on a single subfield. This review arose naturally. We aim to deliver a broad overview of the several competing research lines. We don’t aim for a comprehensive survey (which would not be possible within the reasonable limits of a paper) but rather to illuminate the central questions currently being discussed, allowing curious non-specialists to start exploring the literature on their own. In doing so, we hope to show that the Measurement Problem is not as bleak as it seems, although there is still much to be uncovered.

Given the Measurement Problem's central role, it is unsurprising that several other reviews have been recently published to explore its unresolved aspects from different perspectives. Hance and Hossenfelder \cite{Hance2022} critically assess whether the problem is fundamental or an artifact of our theoretical framework. In a synthetic review, they identify five requirements that any satisfactory solution to the problem must satisfy and analyze several proposed solutions. Müller \cite{Muller2023} followed a different path. Instead of focusing on the proposed solutions, he approaches the Measurement Problem from a formal logic perspective. He decomposes it into six independent problems, which should be investigated separately. Meanwhile, Freire Jr. \cite{FreireJr2024} delivers a historical account, linking the foundations of quantum mechanics to current scientific and technological advances.

Surveys dedicated to objective collapse theories are numerous. Bassi, Dorato, and Ulbricht \cite{Bassi2023} present an in-depth analysis of objective collapse models from theoretical, experimental, and philosophical perspectives. Carlesso \textit{et al.} \cite{Carlesso_2022} further explore experimental tests beyond interferometric setups, while Donadi and Bassi \cite{Donadi2022a} review gravity-induced collapse models. These studies update the comprehensive review by Bassi and collaborators from 2013 \cite{Bassi2013} and build on the foundational report on dynamical reduction models by Ghirardi and Bassi \cite{Bassi_2003}, which remains a central reference in the field. 

Each of these works is worth the attention of readers interested in the status of foundational discussions of quantum mechanics. Nevertheless, none delivers a broad coverage of the Measurement Problem; that’s precisely what we missed when we started our journey. Addressing this gap, our review organizes the various proposed solutions into a structured overview, offering a broad reference point for both specialists and newcomers. 

Our work also brings a distinct perspective by examining the Measurement Problem through the chemistry lens (since two authors, RSM and MB, are theoretical chemists). Historically, chemists have treated quantum mechanics pragmatically, using it as a tool rather than engaging with its foundational questions. However, recent advances in quantum materials \cite{Keimer2017,Campaioli2024}, quantum dots for imaging and drug delivery \cite{Biswas2021}, ultrafast spectroscopy \cite{Zigmantas2022}, and quantum computing \cite{Claudino2022,McArdler2020} have brought the Measurement Problem into sharper focus at the chemistry labs. These fields increasingly rely on quantum coherence, entanglement, and decoherence---concepts central to quantum measurement theory \cite{Schultz2024,Scholes2025}. As chemists push the boundaries of quantum state control, they may need to reconsider long-standing assumptions about wave function collapse, observer effects, and the role of decoherence in chemical processes. 

Despite being unconventional in scientific papers, we extensively referred to non-peer-reviewed web resources, including books, essays, and videos. These resources offer a plain language introduction and analysis free from intimidating technicalities. They naturally don’t replace technical literature, but should not be neglected as relevant entry points to the subject.

\section{Quantum State Description and Evolution}\label{QUANT}

This section briefly introduces basic quantum mechanics concepts, focusing on the elements most relevant to the following discussions. For in-depth treatments, we refer the reader to standard textbooks \cite{Cohen-Tannoudji2020,Sakurai1994}. Concise and accessible overviews can be found in Refs.  \cite{Manzano2020,Wu2024}. 

Quantum mechanics encodes all information about a physical system in the quantum state $\ket\psi$, a complex-valued unit vector in the Hilbert space $\mathcal{H}$. In a finite Hilbert space, all unit vectors are possible physical states of the system.

\begin{figure}[ht!]
    \centering
    \includegraphics[width=0.5\linewidth]{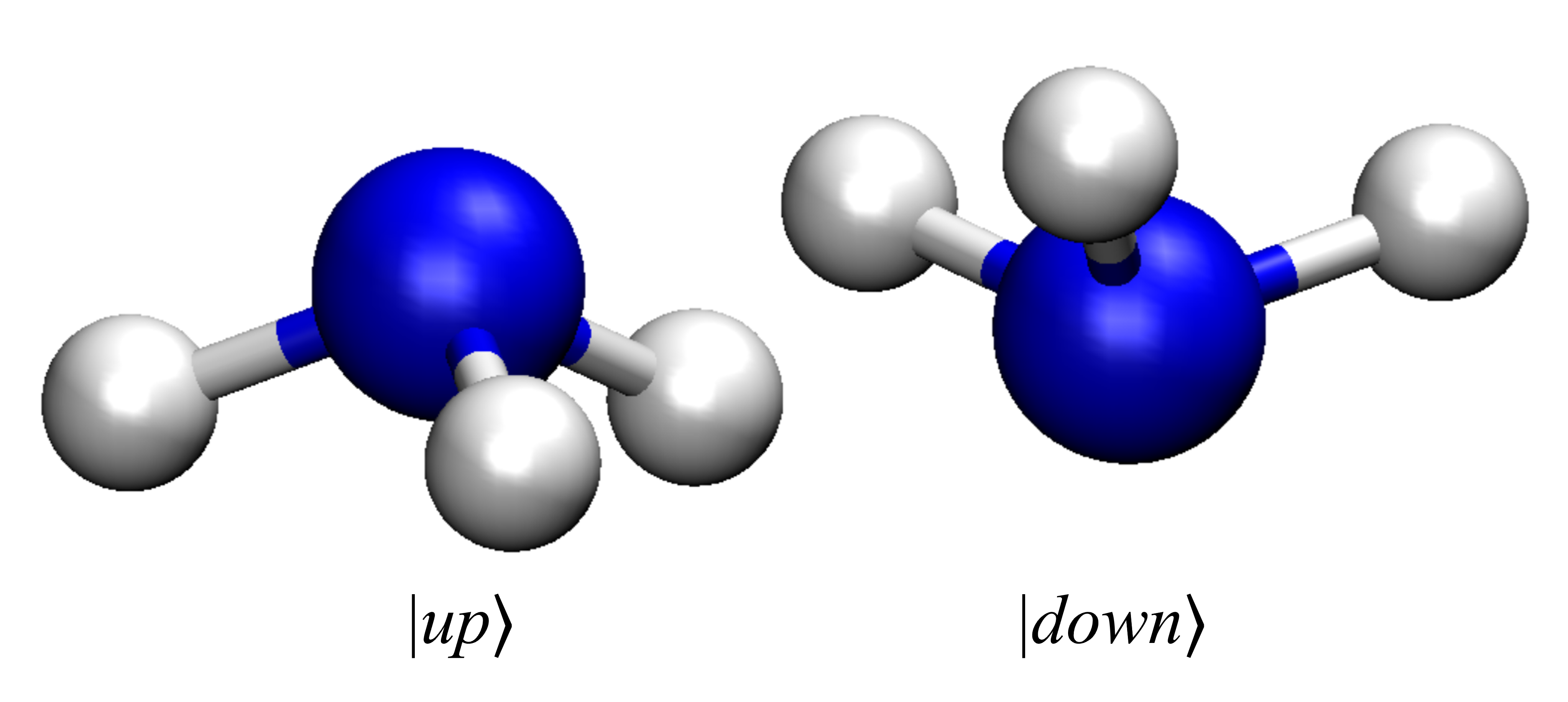}
    \caption{\fontfamily{lmss}\selectfont The ammonia (NH$_3$) molecule serves as an example of a two-state quantum system. Its equilibrium geometry is pyramidal, with the nitrogen atom lying above or below the hydrogen plane. These two geometric configurations, labeled $\ket{up}$ and $\ket{down}$, may form a quantum superposition. }
    \label{fig1}
\end{figure}

Consider, for example, the ammonia molecule NH$_3$ (Figure~\ref{fig1}). Its more stable geometry resembles a pyramid, with the three hydrogen atoms forming a triangular basis and the nitrogen atom lying above or below it. Thus, ignoring molecular vibrations and rotations, the nuclear quantum state of ammonia $\ket{\psi}$ has two components, corresponding to pyramids up and down \cite[Chs 8 \& 9]{Feynman2011}. Let’s name them $\ket{up}$ and $\ket{down}$ states, respectively, meaning we will treat NH$_3$ as a quantum bit (or a qubit).

\begin{figure}[ht!]
    \centering
    \includegraphics[width=0.8\linewidth]{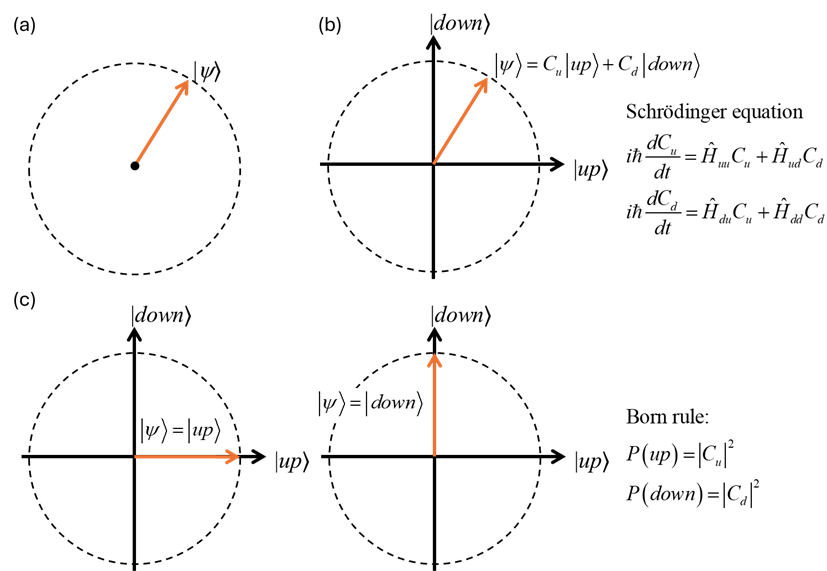}

\caption{\fontfamily{lmss}\selectfont Schematic representation of the quantum description of a two-state system, such as the NH$_3$ molecule in Figure~\ref{fig1}. (a) The quantum state $\ket\psi$ encapsulates all the system’s information and exists as a vector in Hilbert space (illustrated here as a dashed circle plus the central dot). (b) Once a basis is selected, $\ket\psi$ is expressed in terms of its basis components, with complex-valued amplitudes $C_i$. The Hamiltonian operator $\hat{H}$ governs the state’s deterministic evolution via the Schr\"odinger equation. (c) Upon measurement, the state $\ket\psi$ collapses to one of the basis vectors, with the probability of each outcome given by $\vert C_i\vert^2$ according to the Born rule.}
    \label{fig2}
\end{figure}

In this example, the general quantum state $\ket\psi$ is a two-dimensional vector (see Figure~\ref{fig2}-a) written as
\begin{equation}\label{eqQUANT:state}
    \ket\psi = C_u\ket{up} + C_d\ket{down},
\end{equation}
where $C_u$ and $C_d$ are complex-valued coefficients (or amplitudes) with $\vert C_u\vert^2+\vert C_d\vert^2=1$~. Eq.\eqref{eqQUANT:state} trivially generalizes to any number of dimensions as 
\begin{equation}
    \ket\psi=\sum_n C_n\ket{\phi_n}.
\end{equation}
For a continuous basis, it becomes $\ket\psi=\int ds \phi(s)\ket{s}$, and the amplitudes are called \textit{wave functions}.

So far, our example has been restricted to unit vectors, also known as \textit{pure states}. Nonetheless, the usual situation is imperfect knowledge, where we only know that the system is in one of several possible pure states $\ket{\psi_i}$, each occurring with probability $p_i$. In this case, it is more appropriate to describe the system using a density operator 
\begin{equation}
    \rho \equiv \sum_i p_i \ket{\psi_i}\bra{\psi_i},
\end{equation}
which captures statistical uncertainty over an ensemble of pure states. Densities are operators with unit trace and satisfy $\bra{a} \rho \ket{a} \geq 0$ for any normalized state $\ket{a}$, regardless of the chosen basis. A density operator represents a pure state if $\operatorname{Tr}(\rho^2) = 1$; otherwise, the state is \textit{mixed}.

For a composite system consisting of two subsystems $\mathcal{A}$ and $\mathcal{B}$ with Hilbert spaces $\mathcal{H}_A$ and $\mathcal{H}_B$, the total Hilbert space is the tensor product  
\begin{equation}
    \mathcal{H} = \mathcal{H}_A \otimes \mathcal{H}_B.
\end{equation}
A pure state of the total system is a unit vector in $\mathcal{H}$. It is called \textit{separable} if it can be written as  
\begin{equation}
    \ket{\psi} = \ket{a_k} \otimes \ket{b_k};
\end{equation}
otherwise, it is said to be \textit{entangled}.  $\ket{a_i}$ and $\ket{b_i}$ are normalized, not necessarily orthogonal, states of the two systems. A general pure state can be expressed as a superposition of product basis states:
\begin{equation}
    \ket{\psi} = \sum_{i,j} C_{ij} \ket{a_i} \otimes \ket{b_j}.
\end{equation}

For mixed states, a density operator $\rho$ is said to be \textit{separable} if it can be written as
\begin{equation}
    \rho = \sum_k p_k\, \rho_k^{(A)} \otimes \rho_k^{(B)},
\end{equation}
where each $\rho_k^{(A)}$ and $\rho_k^{(B)}$ are density operators on $\mathcal{H}_A$ and $\mathcal{H}_B$, and $p_k \geq 0$, $\sum_k p_k = 1$. Otherwise, $\rho$ is entangled.

The generalization to more than two subsystems is straightforward.

The state vector $\ket{\psi(t)}$ is a dynamical quantity that evolves unitarily according to $\ket{\psi(t)} = U(t,t_0)\,\ket{\psi(t_0)}$, which implies it satisfies the Schr\"odinger equation:  
\begin{equation}\label{eqQUANT:schro}
    i\hbar\frac{d}{dt}\ket{\psi(t)}=\hat{H}\ket{\psi(t)},
\end{equation}  
where $\hat{H}$ is the Hamiltonian operator (see Figure~\ref{fig2}-b). Equivalently, the density operator $\rho(t)$ evolves according to the von Neumann equation:  
\begin{equation}\label{eqQUANT:neumann}
    i\hbar\,\frac{d\rho(t)}{dt} = \left[\hat{H}, \rho(t) \right],
\end{equation} 
where $\left[\hat{H}, \rho \right] = \hat{H}\rho - \rho\hat{H}$ denotes the commutator.

Every measurable physical quantity (observable) is represented by a self-adjoint operator $\hat{O} = \hat{O}^\dagger$ on $\mathcal{H}$. The possible outcomes of a measurement of $\hat{O}$ are its eigenvalues $o_n \in \mathbb{R}$. For a \textit{non-degenerate spectrum}, the probability of obtaining the outcome $o_n$ at measurement time $t'$ is (see Figure~\ref{fig2}-c) 
\begin{equation}
P(o_n) = \left| \braket{o_n | \psi(t')} \right|^2,
\end{equation}  
where $\ket{o_n}$ is the corresponding eigenstate. In terms of the density operator, this probability becomes  
\begin{equation}
P(o_n) = \mathrm{Tr}\left[ \ket{o_n}\bra{o_n} \, \rho(t') \right].
\end{equation}
This formulation corresponds to a \emph{projective measurement} (also known as a \textit{projection-valued measure}, or PVM), where the outcome $o_n$ is associated with the projection operator $|o_n\rangle \langle o_n|$. For more general types of quantum measurements---such as those involving degenerate spectra or \textit{positive operator-valued measures} (POVMs)---we refer the reader to Ref.~\cite{Wu2024} for a pedagogical overview.

According to the standard postulates of quantum mechanics, after the measurement, the system is left in the normalized eigenstate $|o_n\rangle$ if $o_n$ was measured. This state change is referred to as \emph{quantum state collapse}. In this discussion, we assumed that collapse is instantaneous and occurs at $t'$. Standard quantum theory gives no clue as to whether this is precise or not. Recent resonant inelastic X-ray scattering (RIXS) experiments have suggested that state collapse occurs gradually over a finite time \cite{Ignatova2017}. However, this interpretation remains ambiguous, as the observed dynamics could also be explained by progressive decoherence rather than a fundamental modification of quantum measurement. We will return to when and how long it takes for collapse to occur in discussing Objective Collapse Theories in Section \ref{ObCoTh}. The implications of an instantaneous collapse are explored in Section \ref{Measurement}, in the context of quantum field measurements.

\section{The Problems with Quantum Measurements}

To summarize the description of the quantum state evolution discussed in the previous section, standard quantum mechanics prescribes that a quantum state follows two types of time evolution: Left on its own, it evolves with the Schr\"odinger equation, which is unitary and deterministic \cite[Ch 3]{Cohen-Tannoudji2020}\cite[Ch 1]{Sakurai1994}. During a measurement, the system produces a classical outcome with probabilities given by the Born rule. In idealized cases, this is often described by a non-unitary state update (the \textit{projection postulate}), but more generally, a measurement is any transformation that extracts classical information from a quantum system. These processes must be probabilistic and cannot allow communication between distant systems without a physical carrier.

There are different ways to scrutinize this description \cite{Muller2023, Laudisa2022, Maudlin1995}. Here, we follow Schlosshauer \cite[Ch 2]{Schlosshauer2007}, who decomposed it into three problems: the \textit{problem of the preferred basis}, the \textit{problem of the nonobservability of interference}, and the \textit{problem of outcomes}. Each of these challenges arises naturally from the standard formulation of quantum mechanics and highlights a gap in our understanding of what happens when a quantum system is measured. Below, we define these problems in sequence.

\textit{The problem of the preferred basis.} Suppose we have an apparatus to detect whether the NH$_3$ molecule is \textit{up} or \textit{down}. This apparatus is also modeled as a quantum system, initially in a ``ready'' state. If it measures the molecule \textit{up}, its pointer moves to 1; if it detects \textit{down}, it moves to 0 (in the following, we will omit $\otimes$ for brevity):
\begin{equation}
  \begin{aligned}
    &\ket{up}\ket{\text{ready}} \to \ket{up}\ket{1}, \\
    &\ket{down}\ket{\text{ready}} \to \ket{down}\ket{0}.
  \end{aligned}
\end{equation}

If the molecule is initially in the superposition
\begin{equation}
    \ket{\psi} = \frac{1}{\sqrt{2}} \left(\ket{up} + \ket{down} \right),
\end{equation}
the measurement interaction leads to an entangled state:
\begin{equation}\label{outcomeeq}
    \ket{\psi} \ket{\text{ready}} \to \frac{1}{\sqrt{2}} \left(\ket{up}\ket{1} + \ket{down}\ket{0}\right).
\end{equation}
This one-to-one correlation between system and apparatus defines the premeasurement state in the von Neumann measurement scheme~\cite{Schlosshauer2007}, where a collapse (or equivalent mechanism) is needed to produce a definite outcome.

\begin{figure}[ht!]
    \centering
    \includegraphics[width=0.5\linewidth]{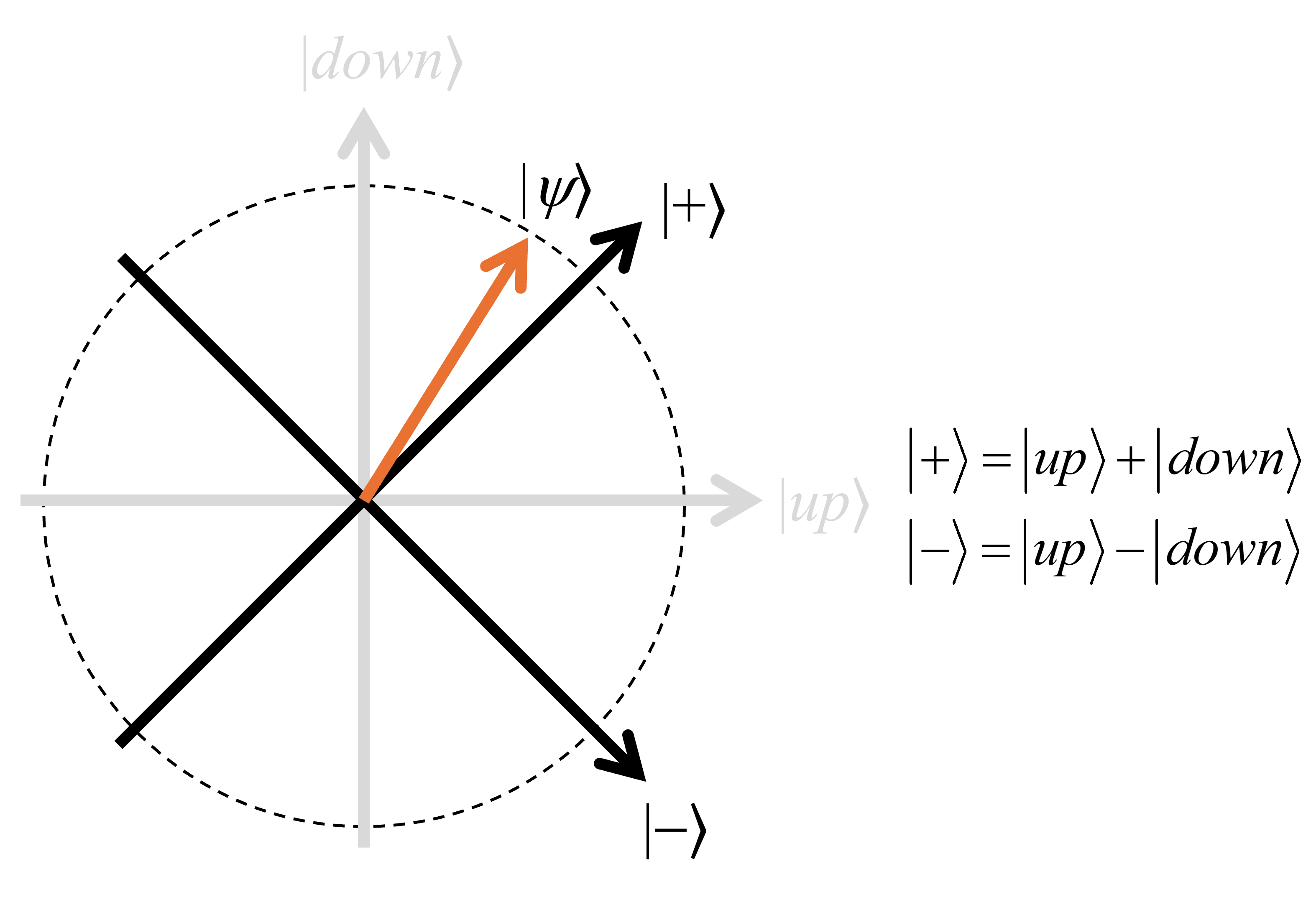}
    \caption{\fontfamily{lmss}\selectfont The choice of basis to describe the quantum system is arbitrary in principle. The quantum state $\ket{\psi}$ can be equally written in terms of the $\{\ket{up},\ket{down}\}$ or $\{\ket{+},\ket{-}\}$ basis. In the premeasurement entangled state, this basis ambiguity extends to the apparatus as well, although only the system’s basis is illustrated here.}
    \label{fig3}
\end{figure}

The same final state can be expressed in a different basis. Define (Figure~\ref{fig3}):
\begin{equation}
  \begin{aligned}
    &\ket{+} = \frac{1}{\sqrt{2}} \left(\ket{up} + \ket{down} \right), \\
    &\ket{-} = \frac{1}{\sqrt{2}} \left(\ket{up} - \ket{down} \right),
  \end{aligned}
\end{equation}
and
\begin{equation}
  \begin{aligned}
    &\ket{A} = \frac{1}{\sqrt{2}} \left(\ket{\text{1}} + \ket{\text{0}} \right), \\
    &\ket{B} = \frac{1}{\sqrt{2}} \left(\ket{\text{1}} - \ket{\text{0}} \right),
  \end{aligned}
\end{equation}
Then the same entangled state becomes
\begin{equation}
    \ket{\Phi} = \frac{1}{\sqrt{2}} \left( \ket{+}\ket{A} + \ket{-}\ket{B} \right).
\end{equation}

Formally, both descriptions are equivalent. But in practice, the apparatus always produces outcomes on a specific basis---here, $\{\ket{1}, \ket{0}\}$---not in $\{\ket{A}, \ket{B}\}$ or any other. One might argue that the apparatus, by design, selects the measurement basis. But this assumes what needs to be explained: how and why a particular basis emerges from the unitary quantum dynamics. After all, the entangled state can be expressed in many incompatible bases, and the Schr\"odinger equation itself does not single out any of them.

This ambiguity---why a specific set of outcomes is realized when many are in principle available---is the \textit{problem of the preferred basis}, first formulated by Zurek~\cite{Zurek1981}.

The \textit{problem of the nonobservability of interference}. Quantum mechanics predicts that systems can exist in coherent superpositions, which should produce observable interference effects. Indeed, such effects are routinely seen at microscopic scales---for instance, in electron double-slit experiments \cite{Tonomura1989} or the oscillations of generalized oscillator strengths for symmetric molecules under electron impact \cite{Barbatti2005}. More recently, Kanitz and coworkers demonstrated interference patterns from helium and hydrogen atoms transmitted through single-layer graphene, achieving diffraction with atoms at kiloelectronvolt energies \cite{Kanitz2024}.

Nevertheless, interference is never observed in macroscopic objects: tables, measurement devices, or Schr\"odinger’s cat do not display interference fringes. One might suspect this absence is simply due to the extremely short de Broglie wavelengths of large systems, making interference effects undetectable. However, this explanation is insufficient. Under carefully controlled conditions, interference remains observable even for massive molecules with minuscule de Broglie wavelengths, as demonstrated by Talbot diffraction experiments with molecules comprising up to 2000 atoms \cite{Fein2019}. Thus, the absence of interference in everyday situations cannot be entirely attributed to technical limitations.

The real puzzle is why coherence, though allowed in principle by the Schr\"odinger equation, becomes unobservable in practice beyond microscopic scales. This disconnect defines the problem of the nonobservability of interference and points toward the need for a dynamical mechanism that suppresses coherence in macroscopic systems.

The \textit{problem of outcomes}. Even after addressing the previous two issues---identifying a preferred basis and suppressing interference---a fundamental question remains: Why do measurements yield a single, definite outcome rather than a superposition of possible results? After decoherence, the total system (including apparatus and environment) is still described by an entangled quantum state encompassing all possible outcomes (as shown in Eq.\ref{outcomeeq}. Yet we never observe such superpositions: each measurement yields a unique result. Indeed, what would it even mean to observe a superposition?

Standard quantum mechanics provides no mechanism for this selection. The Born rule accurately predicts the statistical distribution of outcomes over many measurements. Still, it does not explain why a particular result occurs or what selects it in an individual run.

These three problems define the core of the Measurement Problem \cite[Ch~2]{Schlosshauer2007}. The following section explores how decoherence addresses the first two problems, while a dedicated discussion on the problem of outcomes follows in Section~\ref{FPOUT}.

\section{Decoherence}\label{DECOH}

The problem of the nonobservability of interference is addressed by decoherence. Decoherence \href{https://www.youtube.com/watch?v=igsuIuI_HAQ}{disperses} the quantum information about a system into correlations with the surrounding environment, suppressing the observable effects of superposition \cite{Schlosshauer2007, Ball2017}. To have a feeling of how decoherence acts, suppose that our $\ket{up}$ and $\ket{down}$ system can interact with an environment with only three states, $\ket{e_0}$, $\ket{e_1}$, and $\ket{e_2}$. We initially prepare the system in a superposition of $\ket{up}$ and $\ket{down}$ and the environment in $\ket{e_0}$, such that the initial state of the system plus environment is
\begin{equation}\label{eqDECOH:sysenv}
    \ket{\Psi(t_0)}=(a\ket{up}+b\ket{down})\otimes\ket{e_0}.
\end{equation}
Because of the interaction between the system and its environment, the joint quantum state becomes entangled:
\begin{equation}\label{eqDECOH:entangl}
    \ket{\Psi(t_1)}=a\ket{up}\ket{e_1}+b\ket{down}\ket{e_2}.
\end{equation}
We can monitor the system's evolution alone, averaging it over the environment states. This is done by computing the reduced density matrix for the system. This average is obtained by taking the trace over the environment’s states of the full-density operator:
\begin{align}
    \rho_S &= \mathrm{Tr}_E \left [\ket{\Psi}\bra{\Psi}\right ] \nonumber\\
    & =\begin{bmatrix}
        \vert a\vert^ 2 & ab^\ast\braket{e_2\vert e_1}\\
        ba^\ast\braket{e_1\vert e_2} & \vert b\vert^ 2
    \end{bmatrix}
\end{align}
The diagonal terms $\vert a \vert^2$ and $\vert b \vert^2$ are the $\ket{up}$ and $\ket{down}$ \textit{populations}. The off-diagonal terms are the \textit{coherences} and are responsible for quantum interferences.

In general, the environmental states are not orthogonal, and the coherences are non-null. However, as time passes, the environment states tend to become effectively orthogonal because their dynamics are different when the environment interacts with the system's $\ket{up}$ and $\ket{down}$ geometries. Thus, the off-diagonal terms tend to zero,
\begin{equation}\label{eqDECOH:decoh}
    \rho_S = \begin{bmatrix}
        \vert a\vert^ 2 & ab^\ast\braket{e_2\vert e_1}\\
        ba^\ast\braket{e_1\vert e_2} & \vert b\vert^ 2
    \end{bmatrix}
    \xrightarrow{\text{Decoherence}}
    \begin{bmatrix}
        \vert a\vert^ 2 &    0       \\
              0         & \vert b\vert^ 2
    \end{bmatrix},
\end{equation}
because $\braket{e_1\vert e_2}$ dynamically becomes null. This process is called \textit{decoherence}.

In formal terms, decoherence arises from the time propagation of the reduced density matrix, which evolves with a \textit{master equation} of the type \cite[Ch 4]{Schlosshauer2007}.
\begin{equation}\label{eqDECOH:mastereq}
    \frac{d\rho_S(t)}{dt} = \underbrace{-i\hbar\left[ \hat{H}_S , \rho_S(t)\right]}_{\mathrm{Unitary~evolution}} + \underbrace{\hat{D}\left[ \rho_S(t)\right]}_{\text{Decoherence/\allowbreak dissipation}},
\end{equation}
where $\hat{H}_S$ is the system’s Hamiltonian, including the environment’s perturbation. $\hat{D}$ is the term responsible for decoherence to the environment.

Decoherence tends to be extremely fast when considering the environment as a multidimensional system. For example, we can employ the Joos-Zeh model \cite{Joos1985} to estimate the decoherence time of the $\ket{up}$ and $\ket{down}$ superposition in ammonia gas. According to this model, the coherence in Eq.\eqref{eqDECOH:decoh} decays exponentially,
\begin{equation}\label{eqDECOH:decay}
    \rho_S\left(up,down,t\right) = \rho_S\left(up,down,0\right)e^{-\frac{t}{\tau_D}},
\end{equation}
with a decoherence time given as
\begin{equation}\label{eqDECOH:decohtime}
    \tau_D = \frac{1}{\frac{8}{3\hbar^2}\frac{N}{V}(2\pi M)^\frac{1}{2}a^2(\Delta x)^2(k_B T)^\frac{3}{2}}.
\end{equation}
In this equation, $N/V$ is the number density of the gas at temperature $T$. $M$ is the molecular mass. $a$ is the molecule’s size (the molecule is assumed to be a dielectric sphere), and $\Delta x$ is the displacement between the $\ket{up}$ and $\ket{down}$ geometries. Plugging these quantities in Eq.\eqref{eqDECOH:decohtime}, we find that NH$_3$ coherence disappears almost instantaneously within $10^{-19}$ seconds under room temperature and pressure. Although the superposition persists in the combined system \textit{and} environment, its effects become unobservable due to dispersal in the gas.

Although decoherence often occurs in femtoseconds or even sub-femtoseconds, specific molecular systems exhibit surprisingly long-lived coherence effects. For instance, Kaufman \textit{et al.} \cite{Kaufman2023} demonstrated that electronic coherences could survive for hundreds of femtoseconds in molecules with near parallel potential energy surfaces due to reduced dephasing between branched wavepackets (see also \ref{fig_dyn}). Schultz \textit{et al.} \cite{Schultz2024} comprehensively review cases where vibronic coupling and structured environments lead to long coherence lifetimes, including examples from light-harvesting complexes in photosynthesis. 

Babcock \textit{et al.} \cite{Babcock2024} \href{https://youtu.be/dURJpmZlmDY}{reported} superradiant effects in tryptophan networks extending over microtubule-scale architectures, which require long-lived electronic coherence. This is remarkable, since superradiance typically unfolds over hundreds of femtoseconds \cite{Gross1982}. For isolated tryptophan in aqueous solution, electronic decoherence occurs in less than 100~fs \cite{Ajdarzadeh2023}, implying that the protein scaffold must be critical in preserving coherence to allow superradiant emission. Babcock’s result has also attracted media attention (see \href{https://youtu.be/xa2Kpkksf3k}{here} and \href{https://youtu.be/R6G1D2UQ3gg}{here}) because it may support Hameroff-Penrose’s hypothesis on the quantum origin of consciousness (see Section \ref{mind-collapse}).

These findings suggest that molecular architecture can sometimes counteract decoherence, challenging the assumption that coherence always vanishes rapidly. Theory, especially atomistic simulations \cite{Shu2023},  has still to catch up in this field.

In addition to suppressing quantum interferences, decoherence also addresses the problem of the preferred basis. It \href{https://youtu.be/vSnq5Hs3_wI}{leads to} the selection of a stable outcome basis through a process known as \textit{einselection}, which drives the quantum state to the basis least entangled with the environment (this basis is known as the \textit{pointer basis}) \cite{Zurek1981}. This preferred basis depends on the Hamiltonian \cite{Paz1999}. Assuming that the position operator commutes with the system-environment interaction Hamiltonian (usually the case), decoherence tends to select the position eigenstates as the preferred basis as long as the interaction Hamiltonian is more important for the dynamics than the system’s Hamiltonian. If both terms compete, the preferred basis is one close to coherent states, which minimizes the uncertainty in both position and momentum. If the system’s Hamiltonian dominates the dynamics, energy eigenstates are preferred as the basis.

For a molecule in a gas, Paz and Zurek's analysis \cite{Paz1999} implies that the outcome basis of the environment selects for measuring nuclei is approximately position states. On the other hand, energy states emerge as the preferred outcome basis for measuring electrons. Thus, when we initially chose the $\{\ket{up},\ket{down}\}$ basis for describing the quantum state of the ammonia molecule, we did it because it matched our classical intuition of nuclei corresponding to particles with defined positions. After discussing decoherence, we realize that we may have developed such a classic intuition because the environment selects the $\{\ket{up},\ket{down}\}$ basis and suppresses superposition effects, creating a world of particles with well-defined positions.

Nevertheless, einselection may not be the last word in the preferred basis problem. Adil and co-authors have recently shown that the preferred basis is not solely dictated by the system-environment interaction, but also by the natural factorization of the global Hilbert space \cite{Adil2024}. Different factorizations can lead to multiple coexisting sets of pointer states, meaning classical behavior emerges in different ways depending on the structure of the Hamiltonian. These coexisting classical realms challenge the idea that einselection always yields a single classical outcome. They may even reignite the preferred basis problem, but now at the level of Hamiltonian factorization.

Decoherence explains why quantum superpositions become locally invisible; however, it does not explain why multiple observers consistently agree on what they observe. Zurek proposed that this consensus arises through \emph{Quantum Darwinism}~\cite{Zurek2009,Zurek2022}, a process in which the environment acts not as a sink of information but as a communication channel. As the system interacts with its surroundings, information about its pointer states---those least perturbed by decoherence---is redundantly imprinted across many fragments of the environment. Different observers can then independently access these fragments to infer the same system properties without disturbing them. Objectivity, in this view, emerges as a Darwinian selection of information: only the most robust, redundantly recorded observables survive to be classically shared. 

Decoherence is part of the standard quantum mechanics toolkit. It is an experimentally well-documented effect crucial for quantum mechanical applications, such as quantum information \cite{Wu2024}. Regardless of one’s preferred interpretation, the role of decoherence remains essential and cannot be ignored. 

This dominant role of decoherence in the current understanding of quantum mechanical processes does not imply a consensus. Landsmann, for instance, characterizes decoherence as “an unmitigated disaster” \cite[Ch 11]{Landsman2017}. His core criticism is that decoherence rigorously only occurs in an environment with infinite degrees of freedom after infinite interaction time with the system. Kastner reinforces this criticism by noticing that decoherence claims of describing irreversible processes suffer from similar conceptual problems already present in statistical mechanics when explaining the emergence of irreversibility from reversible processes \cite{Kastner2014}.

\section{Facing the Problem of Outcomes}\label{FPOUT}

In the previous section, we saw that decoherence clarifies a significant portion of the Measurement Problem. However, the \textit{problem of outcomes} is still open: After the basis is chosen and quantum superpositions are suppressed, the system remains in a mixture of possible outcomes. Decoherence does not tell how and why only one of these outcomes is measured. At this point, standard interpretation of quantum mechanics postulates an additional dynamic step, the collapse:
\begin{equation}
\rho_S =
\begin{bmatrix}
|a|^2 & ab^\ast \braket{e_2\vert e_1} \\
ba^\ast \braket{e_1\vert e_2} & |b|^2
\end{bmatrix}
\xrightarrow{\text{Decoherence}}
\begin{bmatrix}
|a|^2 & 0 \\
0 & |b|^2
\end{bmatrix}
\begin{array}{c}
\xrightarrow{\text{Collapse}}
\begin{bmatrix}
1 & 0 \\
0 & 0
\end{bmatrix} \\[1em]
\text{or} \\[1em]
\xrightarrow{\text{Collapse}}
\begin{bmatrix}
0 & 0 \\
0 & 1
\end{bmatrix}
\end{array}
\label{eqFPOUT:collapse}
\end{equation}

\begin{figure}[ht!]
    \centering
    \includegraphics[width=0.7\linewidth]{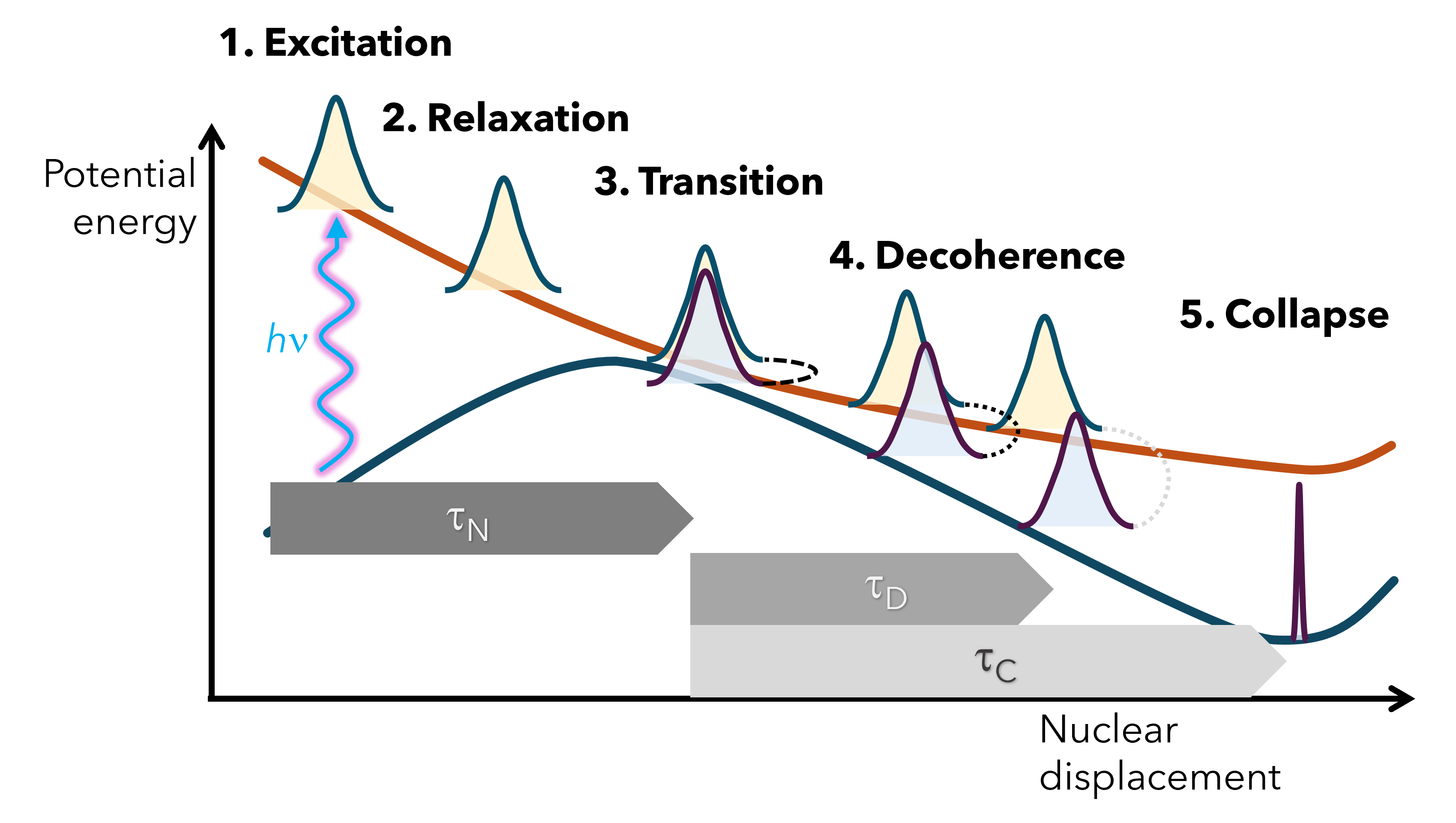}
    \caption{\fontfamily{lmss}\selectfont Schematic illustration of molecular evolution after photoexcitation. The nuclear wavepacket initially relaxes in the excited state until reaching a degeneracy region with the ground state within a time $\tau_N$. There, coherence builds, but it is quickly counteracted by decoherence due to wavepacket dephasing, disappearing within $\tau_D$. Up to this point, the Schr\"odinger equation provides an impeccable description of the phenomenon. If the molecule is measured at $\tau_C$, it will either be in the ground (as illustrated) or in the excited state, with statistics following the Born rule.}
    \label{fig_dyn}
\end{figure}

We illustrate these different processes with an example of a photoexcited molecule undergoing internal conversion in Figure \ref{fig_dyn}. Light initially promotes the molecule to an excited electronic state. The nuclear wavepacket evolves on the potential energy surface of the excited state until it reaches a region of degeneracy with the ground state potential energy surface. There, quantum coherence builds between the two states, creating a state superposition. The dephasing between the wavepacket components in each state counteracts, causing decoherence. All events up to this point, including decoherence, can be simulated by propagating the Schr\"odinger equation for the molecule \cite{Worth2020}. If the molecule is measured, it will be found to be either in the ground or excited state. However, our simulation can only tell the probabilities of these outputs. 

Standard quantum mechanics assumes that a wave function collapse occurred. Nevertheless, the existence of collapse is far from consensual. It is regarded as a superfluous hypothesis \cite{Zeh1993} in several quantum mechanical interpretations.

In the following subsections, we will dive into a few classes of potential solutions that represent the current debate on the Measurement Problem, with little attention to philosophical and historical aspects. More comprehensive approaches can be found in Refs.\cite{FreireJr2024}, \cite[Introduction \& Ch 11]{Landsman2017}, \cite{d'Espagnat2003, Albert2023}.

Before that, we should clarify the concepts we are dealing with. In literature, decoherence and collapse concepts are sometimes used interchangeably. We should avoid it, as these phenomena may, in principle, be completely independent and occur on different time scales. 

As stated in Eq.~\eqref{eqFPOUT:collapse}, decoherence acts on the off-diagonal terms of the system's reduced density matrix. Collapse, however, zeroes all diagonal terms except one in the \textit{entire} density matrix---that is, the matrix describing the system, apparatus, and observer collectively, which Eq.~\eqref{eqFPOUT:collapse} does not represent.
 We can be a bit more flexible with this definition. In principle, a quantum theory including collapse could go as far as to provide a diagonal density matrix. This means we would be left in a situation of statistical uncertainty: the collapse occurred, but the theory only informed us of the probability of each outcome. We could call such a case an \textit{epistemic} collapse. If, however, the theory zeroes the diagonal (except one) elements, then we would have an \textit{ontic} collapse. 

    \subsection{Many-Worlds Interpretation}\label{ManyW}

The lack of information on the outcomes is not a problem for some interpretations. For instance, the Many-Worlds Interpretation (or Relative State Interpretation or, still, Everettian Quantum Mechanics) proposes that the remaining outcomes after decoherence correspond to independent universe branches \cite{Wallace2012}. Observers existing in each branch \href{https://youtu.be/BU8Lg_R2DL0}{perceive} different outcomes because their quantum state, entangled with the molecular quantum state, also branches. In one branch, the combined molecule-observer quantum state may be $\ket{up}$$\ket{observer\mathrm{-}sees\mathrm{-}up}$ and in another, $\ket{down}$$\ket{observer\mathrm{-}sees\mathrm{-}down}$. No observed branches with superpositions of both states exist because of decoherence toward the environment (surrounding the system and the observer) \href{https://youtu.be/GlOwJWJWPUs}{suppressing} them \cite{Zurek2005}. Thus, the Many-Worlds Interpretation posits the existence of a single quantum universe composed of non-communicating classical branches.

For the Many-Worlds Interpretation, the quantum state $\ket\psi$ is a complete description of reality (no hidden variables are required), exclusively evolving according to the Schr\"odinger equation (collapse is only an illusion caused by the universe branching) \cite{Carroll2021, Carroll2019}. Thus, the Many-Worlds Interpretation is consistent with standard quantum mechanics and requires no modification to the theory. However, it remains an interpretation rather than a testable theory, since it currently makes no unique experimental predictions. A significant challenge is the technical infeasibility of maintaining and measuring a quantum system in a coherent superposition of orthogonal states, especially for large systems. This difficulty explains why superpositions of macroscopic states cannot be observed directly due to rapid decoherence. Susskind and Aaronson demonstrated this limitation using methods from quantum computational complexity, naming it the \textit{quantum-hard problem} \cite{Aaronson2020, Susskind2014}. 

A puzzling feature of the Many-Worlds Interpretation is understanding how the deterministic branching relates to the wave function amplitudes and the Born rule. Carroll and Sebens \cite{Carroll2014} propose that this relation is established after decoherence occurs but before the measurement is registered. In this period, if the observer wants to estimate which branch they are, their rational credence directly leads to the Born rule. Thus, the probabilistic character of quantum mechanics would be a subjective aspect arising from the observer's self-locating uncertainty. Carroll and Sebens' derivation of the Born rule is only strictly valid if the ratios of branch amplitudes are the square roots of rational numbers. However, they argue that the large number of branches in practical situations always allows this condition to be approximately satisfied. This approximation also enables experimental tests of their hypothesis by searching for minor deviations from the Born rule. Kent \cite{Kent2015} sharply questioned the conceptual and mathematical soundness of Carroll and Sebens’ account, arguing that their notion of self-locating uncertainty presupposes an objective and well-defined branching of the wave function---something not provided by the Everettian framework itself. 
 
Like Carroll and Sebens, Short \cite{Short2023} derives the Born rule in Many-Worlds Interpretation based on estimating the probability of picking a random branch among many. However, instead of relying on self-locating uncertainty, he does it by establishing three axioms that the probability must satisfy: 1) the probability depends only on the present state (regardless of how it was generated); 2) only branches with non-zero amplitudes count as components of the many-world state; 3) probability cannot flow between uncoupled worlds. Short then shows that the Born rule uniquely satisfies these axioms.

The Many-Worlds Interpretation may seem exotic to non-specialists due to its exuberant ontology. Despite this perception, it is a serious contender among the proposed solutions to the Measurement Problem. It is a popular interpretation among physicists, with prominent supporters like Carroll \cite{Carroll2021, Carroll2019, Carroll2014}, Tegmark \cite{Tegmark2014}, Zeh \cite{Zeh1970,Zeh1993,Byrne2022}, and \href{https://youtu.be/ldgK7EhEnto}{Deutsch}---though their formulations vary in how 
explicitly Everettian they are and in the extent of their commitment to the metaphysics of many worlds. However, it also faces outstanding critics, like Penrose \cite{Penrose2016} and \href{https://youtu.be/AglOFx6eySE}{Albert}. For Landsmann, the Many-Worlds Interpretation would be “acceptable only if truly everything else has failed” \cite[Ch 11]{Landsman2017}.

    \subsection{Epistemic Interpretations}\label{EpistI}

Epistemic interpretations claim that a quantum state is no more than a mathematical tool to predict outcomes. This stance is shared by interpretations such as the Relational Interpretation \cite{Robson2024, Rovelli1996, Rovelli2018}, Quantum Information-Theoretic approaches \cite{Brukner2003}, Quantum Bayesianism \cite{Fuchs2014}, and Indivisible Stochastic Process Theory \cite{Barandes2023}, which are discussed next.

        \subsubsection{Relational Interpretation}\label{RelatI}

The \textit{Relational Interpretation} of quantum mechanics (or \textit{relational quantum mechanics}), proposed by Rovelli, focuses on the interaction between systems, emphasizing that properties are defined only in relation to such interactions \cite{Rovelli2018}. Thus, the molecular geometry may be well-defined for a nearby molecule in the environment acting as an observer while remaining in superposition for another molecule that is too distant to interact significantly.

Rovelli draws attention to the fact that this state dependence on a specific observer is not new in physics. It is analogous, for instance, to the velocity dependence on the particular reference frame in which it is measured. The novelty in quantum mechanics is that the non-commutative nature of quantum variables implies that not all properties can be sharply defined simultaneously \cite{Rovelli2018}, limiting the completeness of information derivable from a single interaction.

In the Relational Interpretation, the wave function is not ontologically real (see Section \ref{ONTOL} for discussion). Instead, it serves as a tool for predicting the probabilities of interaction outcomes. Conceptual issues with measurement arise only when undue ontological weight is assigned to it. By treating the wave function purely as a predictive instrument, measurement can be reframed as an ordinary interaction, resolving these difficulties.

The Relational Interpretation is characterized by a sparse ontology \cite{Rovelli2018}, meaning that properties are not defined between interactions. Nevertheless, it remains a realist interpretation because it posits a physical world of interacting systems. 

Giacomini, Castro-Ruiz, and Brukner have recently extended relational ideas beyond Rovelli's Relational Quantum Mechanics by incorporating \textit{Quantum Reference Frames, QRF} \cite{Giacomini2019}. In their approach, a QRF is simply a quantum system elevated to the role of a reference frame so that its degrees of freedom define the \textit{origin} relative to which one describes other systems. By extending canonical transformations to the quantum domain, Giacomini \emph{et al.} show how usual frame changes (translations, for example) remain valid even if the reference frame itself is in a superposition or entangled state. Their approach leads to the relativity of notions like superposition and entanglement. If a given particle is in a spatial superposition from the lab’s viewpoint, then from the particle’s viewpoint, the lab may itself appear in a corresponding superposition.

This perspective may offer new insights into the Measurement Problem, particularly in Wigner's friend scenarios, where the possibility of conflicting accounts among observers becomes natural if each observer constitutes its own QRF \cite{Giacomini2019, Vanrietvelde2020}. Section \ref{GravObjCol} discusses how Giacomini and Brukner use QRFs to challenge Penrose's conjecture on gravity-induced collapse \cite{Giacomini2022}.

        \subsubsection{Information-Theoretic and QBism Views}

Another epistemic perspective comes from quantum information theory, which treats quantum mechanics as an axiomatic framework akin to thermodynamics. Brukner and Zeilinger \cite{Brukner2003} argue that quantum theory is fundamentally about information, not underlying physical reality. Its postulates describe how information is processed, not how systems evolve in space and time.

Recent experimental results reinforce this perspective. Spegel-Lexne and co-authors demonstrated the equivalence of entropic uncertainty and wave-particle duality \cite{Spegel-Lexne2024}, suggesting that quantum measurement can be understood through informational constraints rather than wave function collapse. 

A more radical variant of this informational approach is \textit{Quantum Bayesianism}, or QBism \cite{Fuchs2014}. QBism interprets the quantum state not as a system property but as a tool that \href{https://www.youtube.com/watch?v=d4-HD8hNmhI}{reflects} an agent's personal beliefs (or more precisely, \textit{doxastic states}) about the outcomes of their measurements. For instance, assigning the state $\ket{up}$ to a molecule expresses an agent’s expectation about what they will experience upon interacting with it, not a claim about an objective feature of the molecule.
When a measurement is made, the quantum state is updated to reflect the agent’s new information, akin to a Bayesian update of probabilities. From this perspective, collapse is not a physical transformation but the agent updating their beliefs in response to their own experience, which is the measurement outcome \cite{Fuchs2014, Franck2017}.

Some no-go theorems are sometimes cited as challenging epistemic views, particularly the Pusey-Barrett-Rudolph (PBR) theorem \cite{Pusey2012}. It shows that, under certain assumptions, quantum states cannot be interpreted merely as statistical knowledge about underlying physical states---that is, hidden variables (see Section \ref{Hidden}). Instead, the quantum state must be part of physical reality. However, this conclusion applies only to models that posit such hidden variables. Interpretations like QBism, which deny the existence of underlying ontic states altogether, fall outside the scope of the theorem \cite{LeiferBlog2011, Leifer2014}. Rovelli has made similar clarifications for relational interpretations as well \cite{Rovelli2025}.

Another supposed challenge to QBism comes from Ozawa’s intersubjectivity theorem \cite{Khrennikov2024ozawa, Khrennikov2024}, which claims that two observers jointly measuring a system must agree on its outcome. This has been interpreted as incompatible with QBism’s agent-centric framework. However, Schack has recently argued that this conclusion is mistaken and that QBism maintains consistency across observers without requiring objective outcomes in the traditional sense \cite{Schack2023}.

A common misunderstanding is to conflate QBism with older subjectivist interpretations, such as that of London and Bauer \cite{London1983}, who suggested that collapse results from an observer becoming aware of an outcome. However, QBism takes a different stance: the measurement outcome is not something the agent discovers; it is the agent’s experience. No underlying \textit{real} wave function remains untouched---it is a theory about actions and consequences, not objective ontologies \cite{debrota2024}.

By rejecting objective quantum states, QBism sidesteps many interpretative paradoxes, offering a consistent but radically subjective view in which quantum theory serves as a decision-making tool for agents.

        \subsubsection{Indivisible Stochastic Process Theory}

Barandes has recently \href{https://www.youtube.com/watch?v=7oWip00iXbo}{proposed} the \textit{Indivisible Stochastic Process Theory}, where classical configurations of particles, fields, or qubits evolve via non-Markovian dynamics \cite{Barandes2023,barandes2023stochasticquantumcorrespondence}. In this type of dynamics, which recovers all results from conventional quantum mechanics, transition probabilities cannot be estimated at intermediate times. In Barandes' approach, wave functions are considered predictive tools only, and the Hilbert space of quantum mechanics is demoted to a convenient instrumentalist framework without an ontological basis.

Barandes \href{https://www.youtube.com/watch?v=sshJyD0aWXg}{draws} a historical analogy, comparing the transition from instrumentalist frameworks in quantum mechanics to the Copernican revolution, where the instrumental Ptolemaic system of epicycles was replaced by a more fundamental description based on forces. Similarly, he advocates for a clear physical picture to supersede the abstract formalism of Hilbert space mechanics. 

Unlike the Relational Interpretation (Section \ref{RelatI}), the Indivisible Stochastic Process formulation posits a non-perspectival, objective reality: each system has a definite configuration that evolves stochastically in time. The ontology is well-defined at all times: each system has a definite configuration, even between interactions. What is sparse are the dynamical laws, since conditional probabilities connecting two moments in time are generally defined only when the earlier time is either the initial time or a special kind of interaction event (called a division event). This framework does not assume divisibility or Markovianity.

By embedding quantum systems within generalized stochastic systems, the Indivisible Stochastic Process formulation aligns with a realist philosophy, reinterpreting quantum systems as governed by dynamic interactions without requiring measurement to be treated as a distinct or special process.

    \subsection{Objective Collapse Theories}\label{ObCoTh}

        \subsubsection{Agnostic Objective Collapse}

Opposed to Many-Worlds Interpretation and epistemic interpretations, \textit{objective collapse theories} consider that collapse is a physical process, and that standard quantum mechanics is still incomplete \cite{Bassi2023, Bassi2013}. Thus, these theories attempt to change the Schr\"odinger equation to account for the collapse. This subsection deals with agnostic objective collapse, that is, theories that do not attribute a cause to the collapse. In the following subsection, we will look at gravity-induced collapse.

\begin{figure}[ht!]
    \centering
    \includegraphics[width=1.0\linewidth]{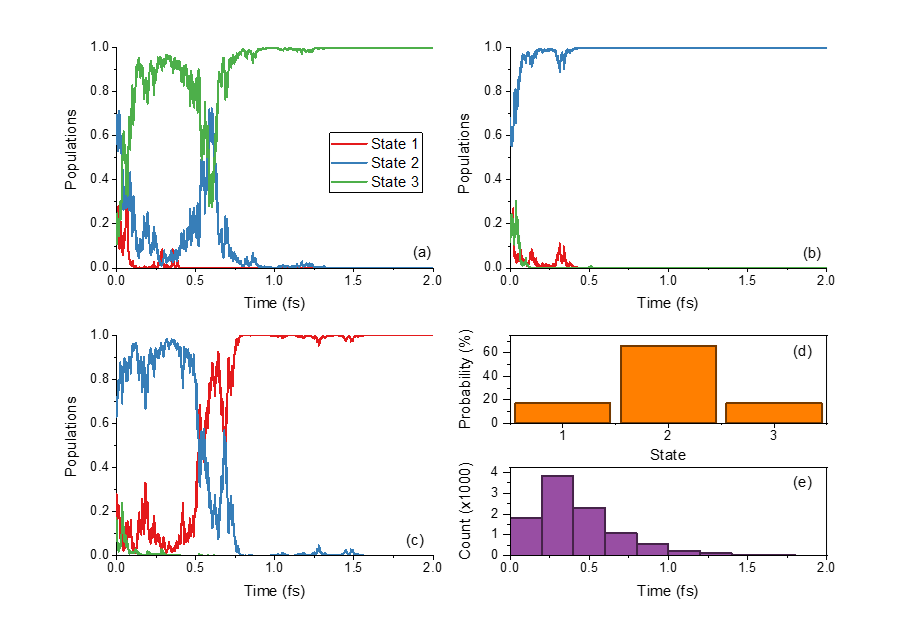}
    \caption{\fontfamily{lmss}\selectfont Example of application of the stochastic Schrödinger equation (Eq.\eqref{eqObjecol:gisin}) taking the Hamiltonian of the one-dimensional harmonic oscillator for the operators $\hat{A}_n$. The system is initially in a superposition $\ket{\psi(0)}=\sqrt{\frac{1}{6}}\ket{1}+\sqrt{\frac{2}{3}}\ket{2}+\sqrt{\frac{1}{6}}\ket{3}$ of the ground and first two excited states. The oscillator’s mass and angular frequency are equal to one atomic unit. $\eta$ was arbitrarily chosen as $0.25~\mathrm{au}$. (a) – (c) Shows the population evolution in three realizations, each collapsing to a different state. (d) The statistics over 10 thousand realizations show that the stochastic formulation recovers the Born rule with $1/6$, $2/3$, and $1/6$ probabilities for states $1$, $2$, and $3$, respectively. (e) For the employed parameters, the collapse occurs within a few femtoseconds. Simulations were performed using the Skitten program developed in our group.}
    \label{fig4}
\end{figure}

Most objective collapse theories predict that a system in quantum superposition randomly localizes in one of the outcomes given enough time \cite{Bassi2013, Adler2012, Ghirardi1990, Torres2024}. The quantum state is driven by a stochastic Schr\"odinger equation, such as the \textit{quantum-state diffusion equation} \cite{Adler2007, Gisin1992}
\begin{equation}\label{eqObjecol:gisin}
d\ket{\psi} = -\frac{i}{\hbar}\hat{H}\ket{\psi}dt + \sum_n \left( \hat{A}_n - \braket{A_n} \right)\ket{\psi}dW_n - \frac{\eta}{2}\sum_n \left( \hat{A}_n - \braket{A_n}\right)^2\ket{\psi}dt.
\end{equation}
In addition to the usual Hamiltonian term, the right-hand side contains two additional terms that tend to localize the state in one of the eigenstates of the operators $\hat{A}_n$, as illustrated in Figure~\ref{fig4} for $\hat{A}_n=\hat{H}$. The collapse is stochastic and controlled by a Wiener process $W_n$ (with $\braket{dW_n}=0$ and $dW_m dW_n=\delta_{mn} \eta dt$), which, for the ensemble density, the statistical distribution of outcomes follows the Born rule. A positive real constant $\eta$ sets the noise strength (diffusion rate), hence the collapse timescale. The equation is nonlinear due to the $\braket{A_n}=\braket{\psi\vert\hat{A}_n\vert\psi}/\braket{\psi\vert\psi}$ terms. However, the evolution of the ensemble density is linear thanks to the quadratic last term \cite{Gisin1989}. This linearity is required to avoid superluminal information exchange.

Different objective collapse theories have been proposed, corresponding to specific choices of operators $\hat{A}_n$, including the Hamiltonian operator \cite{Schack1995}, position operators (\textit{GRW} after the initials of the authors \cite{Ghirardi1986}), density number operators (\textit{CSL} for \textit{Continuous Spontaneous Localization} \cite{Ghirardi1990}), and mass density operators \cite{Diosi1989} (which is discussed in Section \ref{GravObjCol}). These specific models are thoroughly reviewed in Ref. \cite{Bassi2023, Bassi_2003}.

No observers or interacting systems are required in Eq.~\eqref{eqObjecol:gisin}. The localization time \href{https://youtu.be/FP6iyVJ70OU}{depends} on the size of the system. It may take billions of years for an isolated ammonia molecule in superposition to collapse to $\ket{up}$ or $\ket{down}$. In a gas, however, such localization occurs nearly instantaneously.

        \subsubsection{Gravity-Induced Objective Collapse}\label{GravObjCol}

\begin{figure}[ht!]
    \centering
    \includegraphics[width=0.5\linewidth]{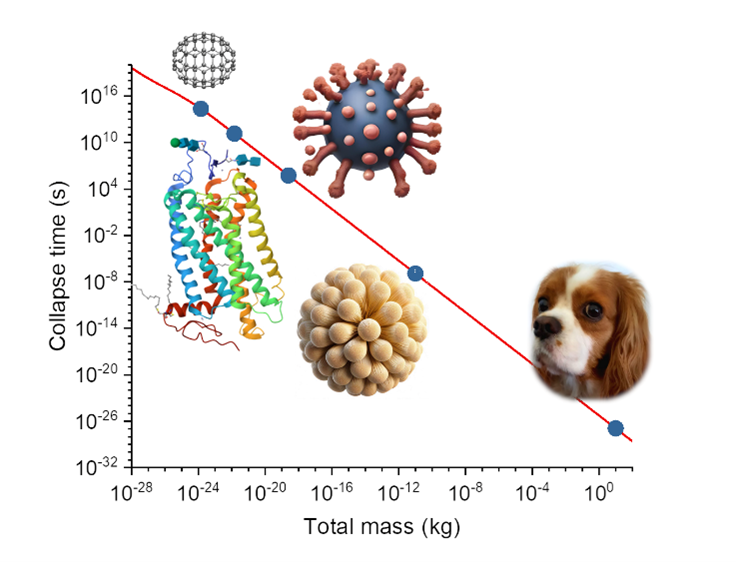}
    \caption{\fontfamily{lmss}\selectfont Di\'osi-Penrose collapse time as a function of the total mass. The estimate was made for a homogeneous carbon system. The collapse times for such a system with the masses of fullerene C$_{70}$, a protein, an adenovirus, a grain of pollen, and a small dog are indicated. Figure reproduced from Ref.~\cite{Tomaz2024}.}
    \label{fig5}
\end{figure}

The highest stakes among the objective collapse theories are those proposing that gravity is the cause of the collapse. Penrose \cite{Penrose1996}, for instance, claims that distinct mass distributions in superposition ($\mu_{up}$ for up or $\mu_{down}$ for down in ammonia molecule) would cause slightly different spacetime distortions. The instability associated with these superimposed distortions leads to the wave function collapse within a time
\begin{equation}\label{eqGIndC:ctime}
    \tau_c = \frac{\hbar}{E_\Delta}
\end{equation}
where $E_\Delta$ is the Newtonian gravitational self-energy
\begin{equation}\label{eqGIndC:selfE}
    E_\Delta = 4\pi G\int d^3\textbf{r} d^3\textbf{r}^{\prime}\frac{\left( \mu_{up}(\textbf{r}^{\prime})-\mu_{down}(\textbf{r}^{\prime}) \right)\left( \mu_{up}(\textbf{r})-\mu_{down}(\textbf{r}) \right)}{\vert \textbf{r}-\textbf{r}^{\prime}\vert},
\end{equation}
and $G$ is the gravitational constant. This collapse time can also be derived by computing the difference between Newtonian free-fall accelerations in the space surrounding each geometry \cite{Howl2019}. A direct application of Eq.~\eqref{eqGIndC:ctime} predicts it would take billions of years to collapse the wave function of an isolated molecule, but only $10^{-27}$ seconds to observe collapse in a $10$ kg body (Figure \ref{fig5}) \cite{Tomaz2024}.

The same collapse time arises from the \textit{Quantum Mechanics with Universal Density Localization} (QMUDL), the model Di\'osi proposed for the stochastic state evolution \cite{Diosi1989}. In QMUDL, the Schr\"odinger equation is modified to
\begin{align}\label{eqGIndC:qmudl}
    d\ket{\psi} = &-\frac{i}{h}\hat{H}\ket{\psi}dt + \int\left( \hat{\mu}(\textbf{r})-\braket{\hat{\mu}(\textbf{r})} \right)dW(\textbf{r})d\textbf{r}\ket{\psi}\nonumber\\
    &-\frac{\kappa G}{4\hbar}\int\int\frac{\left( \hat{\mu}(\textbf{r})-\braket{\hat{\mu}(\textbf{r})} \right)\left( \hat{\mu}(\textbf{r}^{\prime})-\braket{\hat{\mu}(\textbf{r}^{\prime})} \right)}{\vert\textbf{r}-\textbf{r}^{\prime}\vert}d\textbf{r}d\textbf{r}^{\prime}\ket{\psi}dt ,
\end{align}
where the scalar Wiener process is defined by $\braket{dW}=0$ and $dW(\textbf{r})dW(\textbf{r}^{\prime})=\kappa Gdt/\left(2\hbar\vert\textbf{r}-\textbf{r}^{\prime}\vert\right)$. This equation is a particular case of Eq.\eqref{eqObjecol:gisin} where the operators $\hat{A}_n$ are taken proportional to the mass density operator $\hat{\mu}$. The random noise is proportional to the ratio between the gravitational constant $G$ and the reduced Planck constant $\hbar$. $\kappa$ is an arbitrary dimensionless constant. Despite the distinct and independent derivations, the collapse time \eqref{eqGIndC:ctime} is generally called the \textit{Di\'osi-Penrose Model}.

 The Di\'osi-Penrose model has attracted criticism on multiple fronts. Gao \cite{Gao2013} raised several issues on the Penrose conjecture. His main point is that Penrose’s argument by analogy with the conventional quantum mechanical time-uncertainty principle is not rigorous enough. Moreover, he is also right to point out that although we commonly refer to the Di\'osi-Penrose model, Di\'osi and Penrose theories are fundamentally distinct, arriving at the same collapse time through independent, though conceptually related, arguments. Di\'osi himself \cite{Diosi_2022} notes their similarities and differences and suggests that further work is needed to clarify whether the two frameworks are compatible.
 
Giacomini and Bruckner \cite{Giacomini2022} also challenged Penrose’s conjecture that quantum superposition conflicts with Einstein’s equivalence principle. Using the Quantum Reference Frames (QRF) formalism (see also \ref{RelatI}), they showed that it is possible to generalize the principle (or, more precisely, always to make the metric field locally Minkowskian) even in the presence of quantum superpositions of massive bodies.

Additionally, our atomistic simulations showed that the Di\'osi-Penrose collapse time reaches a minimum value independent of superposition displacement and, in some instances, can be surprisingly long \cite{Tomaz2024}. Figurato \textit{et al.} \cite{Figurato_2024} have also independently concluded that the model could have too long collapse times for macroscopic bodies. Beyond these theoretical concerns, experimental tests have also challenged the model. Donadi \textit{et al.} \cite{Donadi2021} and Arnquist \textit{et al.} \cite{Arnquist2022} conducted underground experiments to test gravity-related collapse mechanisms. Their results placed stringent constraints on the model’s parameters (see Section \ref{ExpObjCol}).  

Although all these points must be considered when assessing the validity of the Di\'osi-Penrose model, none of them directly falsifies the model. And, as Thorne said, "Penrose has a way of always being right, even when he does things in a strange way" \cite{Thorne1995}.

Oppenheim has recently proposed another hypothesis that attributes wave function collapse to the influence of gravity. Indeed, his \textit{Postquantum Theory of Classical Gravity} does not focus on the Measurement Problem and has a more ambitious goal of \href{https://youtu.be/yfzosycRoe4}{reconciling} gravity and quantum mechanics \cite{OppenheimPRX2023}. Oppenheim’s central hypothesis is that gravity is a classical field. To ensure classical gravity can interact coherently with quantum matter fields, Oppenheim and colleagues developed a \textit{Classical Quantum (CQ) Dynamics} framework, which circumvents the usual obstacles encountered by semiclassical models, such as violation of the superposition principle due to nonlinearities in the equations of motion \cite{OppenheimPRX2023,Oppenheim_2023,Carlip_2008}.

The CQ dynamics of the hybrid state $\rho(z,t)$ (where $z$ represents the classical phase space coordinates) is expressed in terms of the general master equation \cite{OppenheimPRX2023}
\begin{align}\label{eqGIndC:oppe}
    \frac{\partial\rho}{\partial t} = &-\frac{i}{\hbar}\left[\hat{H},\rho\right] + \sum_{\mu,\nu}\lambda^{\mu\nu}\left( \hat{L}_\mu\rho\hat{L}_\nu^\dagger-\frac{1}{2}\left\{ \hat{L}_\mu^\dagger\hat{L}_\nu,\rho \right\}_+ \right)\nonumber\\
    &+\int dz'\sum_{\mu,\nu}W^{\mu\nu}\left(z\vert z';t\right)\hat{L}_\mu\rho\hat{L}_\nu^\dagger - \frac{1}{2}\sum_{\mu,\nu}W^{\mu\nu}(z;t)\left\{ \hat{L}_\nu^\dagger\hat{L}_\mu,\rho \right\}_+,
\end{align}
which can describe back reactions of quantum matter fields in classical (including gravitational) fields. This equation has completely positive dynamics (hence probabilities remain positive), preserves the trace (probabilities are conserved), and is linear in the density operator (superluminal signaling is avoided). In addition to the commutator term responsible for the unitary evolution and the Lindbladian term \cite{Manzano2020} (proportional to $\lambda^{\mu\nu}$) causing decoherence, the integral term promotes jumps in the quantum and classical subsystems. The last term preserves the norm of the quantum state.

\begin{figure}[ht!]
    \centering
    \includegraphics[width=1.0\linewidth]{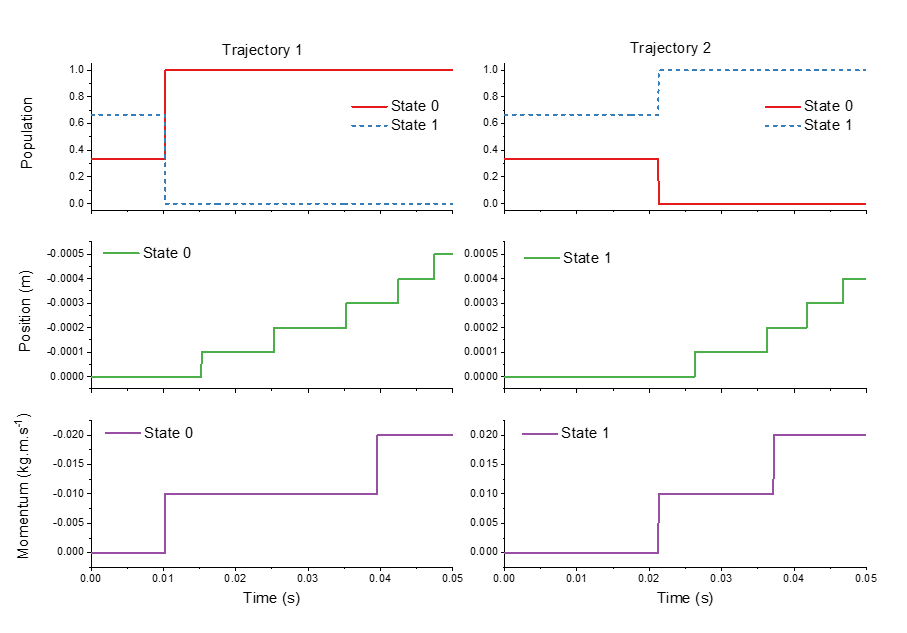}
    \caption{\fontfamily{lmss}\selectfont Evolution of the quantum state population and classical phase space coordinates $(q, p)$ in the CQ dynamics (Eq.\eqref{eqGIndC:oppe} with diagonal Lindblad operators of a qubit interacting with a classical particle \cite{OppenheimQuantum2023}. The initial state is $\delta(q)\delta(p)\left(\sqrt{\frac{1}{3}}\ket{0}+\sqrt{\frac{2}{3}}\ket{1}\right)$. The figure shows two stochastic trajectories. On the left, the interaction between the subsystems collapses the qubit to state $0$, while on the right, it collapses to state $1$. Stochastic jumps in the classical phase space are also observed due to backreaction. If many trajectories are computed, they will be distributed as $1/3$ in $0$ and $2/3$ in $1$. The trajectories were propagated for $0.05$ s, with a timestep of $2.5\times 10^{-5}$ s. The jump rate was $\tau=0.01$~s. The other parameters are $B = 1$~J.s.m$^{-1}$, $m = 1$~kg, and $\omega = 1$~s$^{-1}$. Simulations were performed with the program provided in Ref. \cite{OppenheimQuantum2023}.}
    \label{fig6}
\end{figure}

Like in modeling open quantum systems \cite{Brun2000}, Eq.\eqref{eqGIndC:oppe} is not directly solved for $\rho$ but rather through an ensemble of stochastic trajectories (\textit{unraveling procedure}) \cite{OppenheimQuantum2023, OppenheimNat2023}. In each trajectory, the quantum degrees of freedom undergo stochastic projection onto the eigenstates of the corresponding Lindblad operators, collapsing to a definite outcome in a manner that statistically recovers the Born rule. These quantum jumps induce corresponding phase space jumps in the classical degrees of freedom, representing a backreaction of the quantum system on the classical one. In addition to collapse, this equation also describes decoherence, leading to the suppression of quantum coherence over time. Figure \ref{fig6} illustrates the quantum and classical stochastic jumps in a system composed of a qubit and a classical particle discussed in Ref.\cite{OppenheimQuantum2023}

Even though stochastic trajectories are introduced as a means to propagate the density evolution \eqref{eqGIndC:oppe}, Oppenheim and co-authors assign them ontological significance \cite{OppenheimQuantum2023}. A stochastic trajectory represents the system collapsing to a specific state, just as it is also done in the other objective collapse theories discussed above. Their theory implies that collapse does not require measurement, as it occurs naturally due to interactions with a classical field \cite{OppenheimQuantum2023}. 

Gravity-induced wave function collapse may have profound implications for physics. If any of these theories are confirmed, it would not only close the Measurement Problem but also imply that gravity \href{https://youtu.be/YnXUuyfPK2A}{plays} a unique role in nature, different from the other fundamental interactions. Moreover, it would also impact future efforts to reconcile general gravity and quantum mechanics, moving toward what Penrose calls the gravitization of quantum mechanics \cite{Penrose2014}.

    \subsubsection{Experimental Assessment of Objective Collapse Theories}\label{ExpObjCol}

Experimental efforts to test objective collapse models, especially CSL, have gained remarkable traction in recent years \cite{Carlesso_2022, Bassi_2003, Arnquist2022, Altamura_2025, Gasbarri_2021}. One of the most direct approaches involves matter-wave interferometry with large molecules or nanoparticles, designed to probe the persistence of quantum coherence at mesoscopic scales \cite{Tomaz2024}. Gasbarri \textit{et al.} argue that the microgravity environment of space, with its long free-fall times and low noise, offers a unique setting for such experiments, enabling both interferometric and non-interferometric tests with unprecedented sensitivity to collapse effects \cite{Gasbarri_2021}.

Beyond interferometry, several non-interferometric techniques have been proposed to test collapse models without requiring explicit superpositions. Carlesso \textit{et al.} reviewed platforms such as optomechanical resonators, ultra-cold atoms, and solid-state systems, where collapse-induced noise may manifest as excess heating or anomalous diffusion \cite{Carlesso_2022}. Other proposals investigate spontaneous X-ray emission from the random motion of charged particles, a distinctive prediction of CSL. These diverse strategies have steadily improved the empirical bounds on collapse parameters, ruling out wide ranges of the originally proposed parameter space for both CSL and Di\'osi-Penrose models.

Recent studies have expanded the scope of experimental constraints. Altamura \textit{et al.} analyzed data from the LISA Pathfinder mission and found that rotational degrees of freedom could provide tighter bounds than translational motion in some scenarios \cite{Altamura_2025}. In a related direction, Howl, Penrose, and Fuentes proposed an experiment using a Bose–Einstein condensate prepared in a spatial superposition to test gravitationally induced collapse \cite{Howl2019}.

As discussed in Section \ref{GravObjCol}, the Di\'osi-Penrose model ties collapse to gravitational self-energy. Figurato \textit{et al.} investigated whether the model predicts rapid enough collapse for macroscopic systems and showed that it does not always do so across the entire parameter range \cite{Figurato_2024}. Meanwhile, an underground experiment at the Gran Sasso Laboratory searched for spontaneous radiation arising from collapse-induced diffusion, placing strong limits on the spatial resolution of mass density---enough to exclude the original parameter-free version of the Di\'osi-Penrose proposal \cite{Donadi2021}.

These efforts signal a shift in the status of objective collapse theories from speculative constructs to empirically testable frameworks. Among them, CSL stands out for its tunable parameters and broad compatibility with experimental platforms, while the Di\'osi-Penrose and Oppenheim models offer more rigid, gravity-motivated alternatives. For example, Angeli \textit{et al.} recently demonstrated that if gravity is both classical and local, as in Oppenheim’s proposal (see Section \ref{GravObjCol}), then it must induce diffusion in quantum systems, a phenomenon potentially observable with large-scale mechanical probes such as torsion pendulums \cite{Angeli:2025ojs}.

Altogether, these developments underscore the increasing interconnection of collapse models with experimental physics. Whether or not they ultimately capture physical reality, they provide concrete benchmarks for testing the limits of quantum theory and may yet point toward deeper structures beyond it.

Despite the increasing experimental scrutiny of objective-collapse models, the idea that the Schr\"odinger equation may provide only an incomplete description of physical evolution continues to inspire new proposals. A recent one by Dick reframes the Measurement Problem as an \textit{interaction problem} \cite{Dick2024}. In this view, every inelastic energy exchange between systems acts as a mutual observation, interrupting the smooth Schr\"odinger evolution and producing definite outcomes. The wave function is interpreted as an epistemic superposition of possible ontic outcomes---elastic channels describing continuous evolution, inelastic ones involving discontinuous quantum jumps. Although this approach introduces no explicit modification of the Schr\"odinger equation, it assumes that the equation captures only the evolution of relative probability amplitudes, not the actual transitions that occur during energy exchange. In this sense, Dick’s proposal can be regarded as a minimalist form of objective collapse, locating the breakdown of unitarity within ordinary interactions rather than in new stochastic dynamics. 

    \subsection{Hidden-Variables Theories}\label{Hidden}

        \subsubsection{Physical Restrictions on Hidden Variables}

\textit{Hidden variable theories} (also known as \textit{ontic} theories) \href{https://youtu.be/RlXdsyctD50}{explore} the hypothesis that the quantum state $\ket{\psi}$ may not be a complete description of the system and that a set of additional (hidden) variables may contain further information (accessible or not) \cite{Ghirardi2013}. The hidden variables discussion is where the Measurement Problem spills the most on other fundamental questions of quantum mechanics, such as nonlocality in entangled systems, the completeness of quantum mechanics, and the existence of properties independently of measurement.

Hidden variable theories are typically framed within the limits of two no-go theorems and a few physically motivated restrictions. The theorems are the Bell and the Kochen-Specker ones. Bell’s theorem \href{https://www.youtube.com/watch?v=zcqZHYo7ONs}{states} that no local hidden-variable theory can be entirely compatible with quantum mechanics \cite[Ch 2]{Bell2004} \cite{Mermin1993}. Local in this context means that “the result of a measurement on one system be unaffected by operations on a distant system with which it has interacted in the past" \cite[Ch~2]{Bell2004}.

In turn, the \textit{Kochen-Specker theorem} \cite{Mermin1993, Held2022, Kochen1967} states that any hidden variable theory consistent with quantum mechanics must renounce either all observables having definite values at all times or these values being independent of how they are measured. The theorem is strictly valid for Hilbert spaces with dimension greater than two.

Finally, the physical restrictions are the non-signaling principle (imposing that nonlocality cannot be harvested to exchange superluminal information) and the statistical independence (supposing that initial statistical setups can be completely uncorrelated; see also Section \ref{Deterministic}).

The situation seems simple from there.  For instance, the experimental violations of Bell’s inequalities \cite{Rauch2018}---confirming the predictions of quantum mechanics---imply that physical theories must either be non-local or not have properties defined at all times.  However, no-go theorems are as strong as their underlying hypotheses, often deeply hidden in the deduction. The restrictions imposed by these theorems and physical principles have been constantly challenged \cite{Oaknin2023, Bong2020} (except for the non-signaling principle, which is likely the only consensual rule in the game). Relaxing statistical independence (sometimes loosely called the \textit{free-will principle}) in Bell’s theorem, for example, may allow for building consistent local and realist theories. We will return to this point in Section \ref{Deterministic}.

        \subsubsection{Bohmian Mechanics}

Most of the practical development in hidden-variable theories happens in the framework of Bohmian Mechanics, which posits the existence of corpuscles moving in a non-Newtonian way and guided by a wave function \cite{Dürr2009}. The corpuscle’s velocities are determined by a velocity field (the de Broglie-Bohm Equation) dependent on the wave function, which, in turn, follows the Schr\"odinger evolution. This deterministic time evolution in Bohmian Mechanics automatically solves the problem of outcomes. Although Bohmian Mechanics' experimental predictions are equivalent to quantum mechanics \cite{Das2023}, it is not discarded that they could be experimentally tested \cite{Valentini2010}. A striking example is the experiment by Sharoglazova \textit{et al.} \cite{RN159}, who investigated the energy-speed relationship of wave-guided photons in a microcavity. They observed that for evanescent states within a potential step, lower-energy particles exhibited higher speeds. This result is in tension with standard Bohmian Mechanics, which predicts zero speed for such particles.

Poirier \href{https://quantum-dynamics-hub.github.io/VISTA/77-episode/index.html}{has proposed} an alternative theory that prescribes non-Newtonian trajectories but entirely discards wave functions \cite{Poirier2010, Poirier2020}. In this case, the quantum system is described by an ensemble of interacting deterministic real-valued trajectories with well-defined positions and momenta. This type of theory inscribes itself into the tradition of hydrodynamic interpretation of quantum mechanics, inaugurated by Madelung right after Schr\"odinger’s original work \cite{Madelung1927}\cite[p.~222]{Messiah1961}. In Madelung’s interpretation, quantum mechanics is modeled as a quantum fluid \cite{Holland2005}, which can be understood as \textit{many interacting classical worlds} \cite{Hall2014}.

This picture of many interacting classical worlds has been the basis of molecular dynamics simulations, where the nuclear wave function time evolution is commonly approximated by interacting and non-interacting trajectories, depending on the approach \cite{Crespo-Otero2018, Ibele2024}. Tully, who proposed surface hopping, the most popular method for nonadiabatic molecular dynamics simulations, employed the Madelung Hydrodynamic Interpretation to justify using mixed quantum-classical approximations \cite{Tully1998}. These molecular dynamics simulations, however, adopt the trajectory approach pragmatically without any foundational claim.

    \subsection{Deterministic Models}\label{Deterministic}

Bell's theorem shows that no local hidden variable theories can reproduce all quantum mechanical predictions \cite{Bell2004}. Thus, the experimental violation of Bell's inequalities seemed to be the end of any local deterministic model. However, Bell's theorem explicitly requires the validity of the \textit{statistical independence assumption}, which states that the experimental settings are not influenced by any hidden factors determining the system's evolution. This section surveys two local deterministic models---Superdeterminism and Retrocausality---built on the challenge to the statistical independence assumption. 

\textit{Superdeterminism} \href{https://youtu.be/ytyjgIyegDI}{posits} that hidden correlations between measurement settings and system properties predetermine quantum outcomes \cite{Hossenfelder2020}. ’t Hooft proposed a concrete realization of this idea in his Cellular Automaton Interpretation (CAI) \cite{thooft2016}. In this model, the universe is fundamentally described by a discrete, classical system evolving in time, much like a computer program updating the values of a grid according to fixed rules. These underlying configurations, called ontological states, evolve deterministically and form the actual physical reality. The familiar quantum formalism emerges only as a statistical approximation of our ignorance about the precise ontological state. In this view, superpositions are not real: they are computational tools that summarize ensembles of possible classical states. ’t Hooft argues that violations of Bell inequalities do not imply nonlocality, but instead reflect our mistaken identification of such statistical templates with physical systems.

A recent proposal by Donadi and Hossenfelder \cite{Donadi2022} introduced a local, deterministic model for wave function collapse that explicitly violates statistical independence while reproducing standard quantum predictions. Their model suggests that all quantum states, except measurement eigenstates, are unstable under hidden-variable perturbations and deterministically collapse before reaching the detector. Unlike objective collapse models (see Section \ref{ObCoTh}), the collapse dynamics in the Donadi-Hossenfelder model are entirely deterministic. 

\textit{Cosmic Bell Experiments} have provided stringent tests of the statistical independence assumption in Bell's theorem \cite{Rauch2018,Handsteiner2017}. These experiments avoid the possibility that hidden variables could influence detector settings by randomly determining measurement choices using distant astrophysical sources. By doing so, they attempt to close the \textit{freedom-of-choice} loophole, which is crucial for evaluating the plausibility of superdeterministic explanations. In a landmark experiment, Rauch \textit{et al.} \cite{Rauch2018} used high-redshift quasars---whose light originated billions of years ago---to set measurement parameters in a Bell test. Their results remained consistent with standard quantum mechanics, implying that superdeterministic correlations, if they exist, should date back to the early universe. Similarly, Handsteiner et al. \cite{Handsteiner2017} employed light from distant galaxies to eliminate potential local influences on detector choices. While these results do not \textit{strictly} rule out superdeterminism, it must account for cosmological-scale correlations to remain viable, raising deep challenges for its plausibility.

\textit{Retrocausal Interpretations} also challenge the statistical independence assumption, but, even bolder than Superdeterminism, it assumes influence from future events. Such interpretations propose that the Measurement Problem arises from assuming a one-way flow of time. Models such as the \textit{Two-State Vector Formalism} \cite{Aharonov1964} describe quantum systems by an initial state evolving forward in time and a final state evolving backward from the measurement outcome. The interaction of these forward- and backward-evolving states determines the probabilities of different results. The Pechukas' forces \cite{Pechukas1969}, which are still the primary justification used today for momentum rescaling along nonadiabatic coupling vectors in molecular dynamics \cite{Toldo2024}, were derived from this same type of forward and backward propagation to determine quantum transitions in atomic collisions.

Retrocausality is still an active research field. Adlam \cite{Adlam2022} proposed that accepting nonlocality and imposing relativistic constraints (such as the nonexistence of a preferred reference frame) leads to retrocausality. Sutherland \cite{Sutherland2022} derived the Born rule within a retrocausal model. This is a significant result because it changes the status of the rule from a postulate to a natural consequence of incorporating time-symmetric boundary conditions in quantum mechanics. A notable development is the \textit{Fixed-Point Formulation} proposed by Ridley and Adlam, which presents an atemporal, all-at-once framework for retrocausality and includes a full derivation of the Born rule \cite{Ridley2023,Ridley2024}. Another recent contribution is van der Pals’ \textit{retrocausal hidden-variable model} \cite{vanderPals2024}, which attributes the emergence of definite outcomes to resonance conditions between advanced and retarded virtual photons. These photons arise from an underlying, time-symmetric periodic process occurring beneath the Heisenberg uncertainty threshold.

Both Superdeterminism and Retrocausality remain controversial and experimentally unverified. Nevertheless, despite their speculative nature, these models provide a striking contrast to interpretations that rely on wave function collapse, branching universes, or fundamental randomness \cite{Gisin2021}, keeping the possibility of a fully local and deterministic resolution to the Measurement Problem open.

    \subsection{Dualist Collapse Hypotheses}\label{Dualist}

        \subsubsection{Classical-Apparatus Inducing Collapse}

The core of the \textit{Copenhagen Interpretation} of quantum mechanics, developed by Bohr and Heisenberg, is the assumption that any experiment in physics must be described in terms of classical physical concepts \cite[Introduction]{Landsman2017}. When such an experiment falls within the realm of quantum mechanics, the tension between the system's quantum description and the apparatus's classical description is the crucial factor that causes statistical uncertainty and ultimately selects a single output \cite{Bacciagaluppi2009}. There is no claim that the classical apparatus is not composed of quantum particles. It is just that the description of the experiment is restricted to the use of classical terminology. 

This quantum-classical duality, formulated as the \textit{Heisenberg Cut} arbitrarily separating classical from quantum systems, still plays a role in a pragmatic definition of measurement \cite{Grimmer2023}. It is also at the basis of the \textit{asymptotic emergence hypothesis} \cite{Landsman2013}, which poses that collapse is a unitary physical process. Asymptotic emergence employs algebraic quantum theory to demonstrate that some phenomena forbidden in the quantum limit (like collapse) are allowed in the classical limit due to some exponential sensitivity to perturbations in the Hamiltonian \cite{Landsman2013}.

In a series of works \cite{Schonfeld2022, Schonfeld2023}, Schonfeld has explored cloud chambers, Geiger counters, and Stern-Gerlach experiments for empirical tests of quantum measurement, including possible deviations from the Born rule in detecting rare events. He challenges the conventional view that measurement is an axiomatic feature of quantum mechanics, proposing that measurement emerges phenomenologically from collective interactions in quantum detection systems. He analyzed the detailed microstructure of real measurement systems, using classical idealizations selectively where it seems mathematically or intuitively reasonable.

Oppenheim’s Postquantum Theory provides another perspective for the classical-quantum duality \cite{OppenheimPRX2023}, treating gravity as a classical entity interacting dynamically with quantum systems, causing wave function collapse (see Section \ref{GravObjCol} for a detailed discussion).

        \subsubsection{Mind-Inducing Collapse}\label{mind-collapse}

A historically relevant but discredited hypothesis is that consciousness causes collapse. Wigner advocated for it until he learned about Zeh’s work on what would come to be known as decoherence \cite{Mehra1995}. We mainly mention this mind-matter duality here to disentangle a few ideas that are sometimes mixed up.

First, there is the actual proposal that consciousness causes the collapse, which presupposes a matter-mind duality. The discomfort this hypothesis raises in our physical understanding of the world has never been better embodied than in Bell’s quip \cite{Bell1990}, “Was the wave function of the world waiting to jump for thousands of millions of years until a single-celled living creature appeared? Or did it have to wait a little longer, for some better qualified system ... with a PhD?”

Then, there is an epistemic proposal where the mind does not objectively cause the collapse. Instead, the apparent collapse is an illusion arising when the observer becomes aware of the measurement outcome. We find such a perspective in London and Bauer \cite[p.~251]{London1983} who, in 1939, attributed the essential role played by consciousness to be “the increase of knowledge, acquired by the observation.” 

Finally, Hameroff and Penrose \cite{Hameroff2014} \href{https://youtu.be/xa2Kpkksf3k}{put} forward precisely the opposite hypothesis: collapse causes consciousness. (And quantum coherences in the microtubules mentioned in Section \ref{DECOH} would be part of this process.) They argue that if consciousness cannot be reduced to an algorithmic process, then its physical origin must be in the stochastic nature of the wave function collapse, which, for Penrose, is an objective event, as discussed in Section \ref{ObCoTh}.  

    \subsection{Measurement in Quantum Field Theory}\label{Measurement}

Quantum field theory (QFT) poses specific challenges to the Measurement Problem that are more daunting than in standard, non-relativistic quantum mechanics \cite{Fewster2023}. Because QFT integrates quantum mechanics and special relativity, locality is a built-in concept. This differs entirely from non-relativistic quantum mechanics, where locality is an add-on assumption, like when Bell imposed it in deriving his theorem. Sorkin showed that applying standard collapse rules to QFT can lead to \textit{superluminal information transfer} in scenarios he termed \textit{impossible measurements} \cite{Sorkin1993}. More recently, this issue has been reframed as a violation of the broader \textit{no-signaling} principle, which prohibits communication without a physical carrier \cite{Gisin2024}.

\begin{figure}[ht!]
    \centering
    \includegraphics[width=0.5\linewidth]{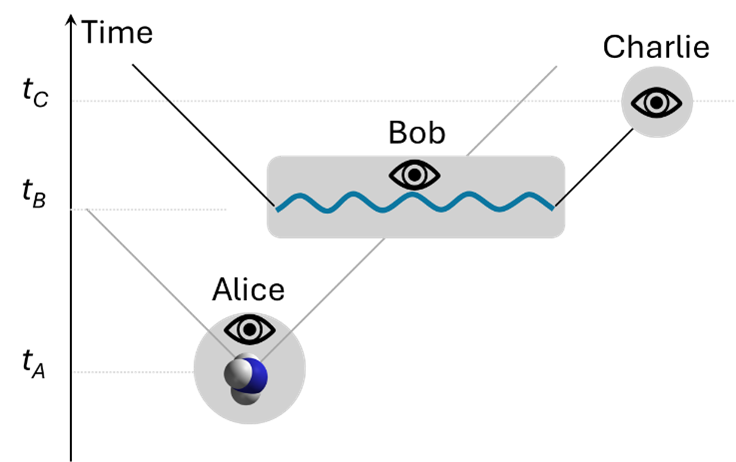}
    \caption{\fontfamily{lmss}\selectfont Impossible measurement in QFT adapted from Ref.\cite{Sorkin1993} Charlie, at time $t_C$, cannot know if Alice flipped the molecule at $t_A$ or not. Could an instantaneous collapse of the electromagnetic field’s state in Bob’s measurement at $t_B$ enable superluminal information transfer between Alice and Charlie?}
    \label{fig7}
\end{figure}

Figure \ref{fig7} illustrates a version of Sorkin’s impossible measurement using our NH$_3$ qubit. Alice, Bob, and Charlie are ready to perform measurements on a molecule-field coupled system at times $t_A$ (Alice), $t_B$ (Bob), and $t_C$ (Charlie). They are spaced to allow causal effects between Bob and Alice and between Charlie and Bob, but not between Alice and Charlie. In the relativity lingo, Alice and Charlie are \textit{spacelike} separated.

Alice prepares a molecule in a $\ket{down}$ state and can choose whether to flip it to $\ket{up}$ at time $t_A$ or not. The molecule is coupled to an electromagnetic field. The field state is $\ket{f_u}$ if the molecule is $\ket{up}$ and $\ket{f_d}$ if the molecule is $\ket{down}$. The field is initially $\ket{f_d}$ and, after $t_A$, it is updated according to Alice’s choice. At $t_B$, Bob measures the field projected on $\left(\ket{up}\ket{f_u}+\ket{down}\ket{f_d}\right)/\sqrt{2}$. Charlie, who, at time $t_C$, has no direct information about Alice’s choice, attempts to indirectly discover that by measuring the field at $t_C$, which was updated after Bob’s intervention. Sorkin argues that, in principle, Charlie's measurements on the field could reveal Alice’s choice, implying superluminal exchange. Therefore, avoiding such impossible measurements would require nontrivial constraints on which observables are consistent with causality.

Such imprecision in the description of the measurement in QFT has been characterized, with a flair to the dramatic, as “a major scandal in the foundations of quantum physics” \cite{Earman2014}. Recently, Bostelmann, Fewster, and Ruep \href{https://www.youtube.com/live/RSn1gaS2mok}{proposed} that the situation is naturally solved within algebraic QFT as long as probes and couplings are constrained to be local \cite{Fewster:2018qbm, Fewster2023, Bostelmann2021}.

This approach, originally developed as a general framework for describing measurement in algebraic QFT, was not tailored to solve the Sorkin problem, but rather aims to identify and localize induced system observables and to characterize state updates. The application to the Sorkin scenario, as presented in \cite{Bostelmann2021}, thus constitutes a nontrivial test of the framework. Their analysis is valid for general fields and spacetime geometries, making it a promising step toward a rigorous measurement theory in QFT. 

A complementary approach is proposed by Polo-G\'omez, Garay, and Mart\'in-Mart\'inez, who develop a measurement theory for QFT based on spatially smeared particle detector models---such as the Unruh-DeWitt detector---that interact locally and covariantly with quantum fields \cite{Polo-Gomez2022}. Their framework, applicable to general fields and spacetime geometries, avoids problems such as Sorkin’s impossible measurements and allows localized interactions without invoking instantaneous collapse. While Bostelmann \textit{et al.} pursue a mathematically rigorous algebraic route, Polo-G\'omez \textit{et al.} adopt a more operational strategy grounded in how detectors realistically interact with quantum fields.

Other proposals inspired by quantum field theory and signal analysis take a different route, using mathematical tools to reinterpret collapse as a statistical construction. For instance, Morgan \href{https://youtu.be/61H0o8W9xg8}{proposed} a framework based on classical random fields and operator-algebraic methods to model quantum phenomena, offering an alternative strategy that remains within a classical probabilistic setting \cite{Morgan:2021jeh}. In his work, these fields---equipped with nontrivial correlation structures---are used to reproduce quantum statistics and construct joint probabilities even for noncommuting observables. His emphasis lies not on deriving definite outcomes from quantum theory, but on modeling the empirical structure of datasets generated by repeated measurements. This approach aims to preserve locality and realism, challenging standard assumptions about the necessity of wave function collapse or quantization itself.

\section{The Ontological Status of the Wave Function}\label{ONTOL}

One fundamental question underlying the Measurement Problem is whether the wave function represents an actual physical object ($\Psi$-\textit{ontic}) or is merely a mathematical tool encoding information ($\Psi$-\textit{epistemic}) \cite{Hubert2022}. The Many-Worlds Interpretation (Section \ref{ManyW}), Objective Collapse Theories (Section \ref{ObCoTh}), and Hidden-Variable Theories (Section \ref{Hidden}) align with the $\Psi$-ontic view, while epistemic interpretations (Section \ref{EpistI}) are, by definition, $\Psi$-epistemic.

A common argument for the $\Psi$-epistemic view is that the wave function evolves in an abstract, high-dimensional configuration space, making it unlikely to correspond to anything physically real \cite{Hubert2022}. However, such a criticism does not intimidate $\Psi$-ontic proponents, who argue that the wave function describes something fundamental about reality, even if not in three-dimensional space \cite{Chen2019, Ney2023}. \textit{Wave function realism}, for example, holds that the wave function is a physical field existing in a high-dimensional configuration space \cite{Albert1996}. Carroll and Singh take this high-dimensional $\Psi$-ontic stand to the extreme \cite{Carroll2019}. In an interpretation they call \textit{Mad-Dog Everettianism}, they propose that the only fundamental elements of reality are the Hilbert space, state vectors, and hamiltonians. Everything else---including space, time, and classical variables---emerges from these.

Other $\Psi$-ontic interpretations have been developed to avoid the conceptual challenges of high-dimensional spaces. One approach is the \textit{multi-field interpretation}, which states that an $N$-body wave function is a multi-field that assigns properties to sets of $N$ points in the 3D space \cite{Hubert2018}. This is the view one of us adopted in an essay discussing the nature of the molecular wave function \cite{Barbatti2023}. Gao \cite{Gao2017} proposed an alternative $\Psi$-ontic interpretation in the 3D space that does not require multi-fields. In his view, the wave function represents a \textit{random discontinuous motion} (RDM) of particles in real space. The wave function phase encodes momentum flow, giving it a concrete physical meaning. Moreover, RDM makes testable predictions, suggesting that collapse is an emergent effect of underlying stochastic motion rather than a fundamental postulate. Whether this model extends to relativistic quantum field theory remains an open question.

\textit{Spacetime state realism} reinterprets quantum mechanics by assigning quantum states to regions of spacetime rather than treating the wave function as an object in a high-dimensional space \cite{Wallace2010}. This approach aligns more naturally with quantum field theory, which assigns quantum states to localized spacetime regions. A similar shift in emphasis to local quantities occurs in Density Functional Theory (DFT) \cite{Jones2015}, which, rather than working with the entire many-body wave function, pragmatically describes fermionic systems in terms of their density---a function of three spatial coordinates.

A different approach comes from the \textit{Nomological Interpretation}, which suggests that the wave function is not a physical object but a law of nature governing quantum behavior \cite{Chen2019}. This is a common stance in interpreting Bohmian Mechanics \cite{Dürr2009, Goldstein2001}.

The Berry phase, which may arise during the cyclic evolution of quantum systems \cite{Berny1984, Xiao2010}, is a mathematical feature of the wave function that can be explored to assess its ontological status. Such a phase can build, for instance, when the nuclear geometry of a molecule evolves along a closed path encircling a conical intersection. (A conical intersection is a molecular geometry at which two adiabatic electronic states degenerate, forming a conical topology of the potential energy surfaces in nuclear coordinate space \cite{Domcke2004}.) Unlike dynamic phases, the Berry phase is tied to the wave function’s geometric properties and may, in principle, manifest in observable interference effects. Valahu \textit{et al.} \cite{Vahalu2023} claimed to have observed such geometric-phase interference in a trapped-ion quantum simulator mimicking conical intersection dynamics, which would strongly support the $\Psi$-ontic interpretation. However, it is unclear whether such effects could be observed in actual molecules since their experiment could be considered more of a controlled simulation than a molecular measurement. Moreover, Min \textit{et al.} \cite{Min2014} and, more recently, Ibele \textit{et al.} \cite{Ibele2023} have provided strong theoretical evidence that the Berry phase originates from the Born-Oppenheimer approximation and disappears in an exact electron-nuclear treatment, suggesting it may be an artifact rather than a fundamental quantum property. 

Although the wave function's ontological nature remains unsettled, some recent experiments may support wave function realism. In particular, Costello \textit{et al.} \cite{Costello2021} reconstructed Bloch wave functions in GaAs semiconductors using angle-resolved photoemission data, showing that wave functions can, in some contexts, be experimentally accessed rather than only inferred. However, such reconstructions apply to quasiparticles and rely on a specific preparation context, and do not settle the deeper question of whether the wave function represents an element of reality. As Rovelli has emphasized \cite{Rovelli2025}, a wave function may encode the full information needed to predict measurement outcomes relative to a given observer, without implying a universally ontic status.

\section{Through the Forest of Quantum Foundations}\label{CONCL}

It is easy to get lost in the Measurement Problem. The gargantuan amount of literature, the innumerable distinct theories, the endless debates, and the entanglement between physics, philosophy, and formal logic all lend anyone stepping into this field a feeling analogous to being trapped in a dense tropical forest. This review aims to offer an updated global map, which, although it cannot pinpoint the escape route, as it simply doesn’t exist, will at least help us steer clear of dangerous cliffs.

Here, we provide a general overview of possible solutions to the Measurement Problem, which remains an actively discussed topic among physicists and philosophers, as well as of newly proposed theories that we consider particularly promising or disruptive. They fall into five classes of explanations: Many-Worlds Interpretations, Epistemic Interpretations, Objective Collapse Theories, Hidden-Variable Theories, and Dualistic Collapse Hypotheses.

Today, the Measurement Problem is much less puzzling than it once was. Thanks to the decoherence program, we understand how quantum interferences are suppressed and how the environment selects the outcome basis. Nevertheless, we still lack a fundamental explanation for why only a single outcome is observed in experiments, and we are still taking the first steps toward a well-defined measurement theory in quantum field theory. Indeed, given that QFT is likely the most successful physical theory ever developed, any satisfactory solution for the Measurement Problem must also be an acceptable solution for measurement in QFT. Therefore, it is worrisome to note that researchers on the Measurement Problem do not usually deal with QFT (and vice versa).

Most scientists who use quantum mechanics in their daily work focus on its practical applications rather than its interpretation. Indeed, it makes no difference whether we interpret the result of an experiment as the quantum state collapsing into a single branch or experiencing a branch that cuts off communication with all others. Such a subtle distinction belongs to defining a preferred worldview. It is even less consequential than choosing between Newtonian and Hamiltonian formulations of classical mechanics, which at least differ in their practical use. The exception concerns objective collapse theories, which propose new physics beyond conventional quantum mechanics. In any case, it makes sense to keep challenging all interpretations and theories to prune the weak and reach complete consistency on the remaining, even if we never get a single consensual description of quantum mechanics.

\section*{Acknowledgments}

The authors gratefully acknowledge Stephen Adler, Jacob Barandes, Matteo Carlesso, John deBrota, Lajos Di\'osi, Christopher Fewster, Nicolas Gisin, Peter Morgan, Tzula Propp, Michael Ridley, Rold\~ao da Rocha, Carlo Rovelli, Jonathan Shonfeld, Barbara \v{S}oda, Niklas S\"ulzner, and Mark van der Pals for their careful reading of the first version of the manuscript and for their thoughtful and precise comments, which encouraged us and helped improve this challenging work.

This work received support from the French government under the France 2030 investment plan as part of the Initiative d’Excellence d’Aix-Marseille Université (A*MIDEX AMX-22-REAB-173 and AMX-22-IN1-48) and from the European Research Council (ERC) Advanced Grant SubNano (grant agreement 832237).

\bibliographystyle{unsrtnat}
\bibliography{refs_overview_final}

\end{document}